\documentclass[english,12pt]{revtex4-2}
\usepackage[T1]{fontenc}
\usepackage[latin9]{inputenc}
\usepackage{color}
\usepackage{array}
\usepackage{float}
\usepackage{booktabs}
\usepackage{multirow}
\usepackage{amsmath}
\usepackage{graphicx}
\usepackage{subscript}

\makeatletter

\providecommand{\tabularnewline}{\\}
\newcommand{\lyxdot}{.}

\makeatother

\usepackage{babel}
\begin{document}
\title{Tuning the electronic and optical properties of $hg-C_{3}N_{4}$ quantum
dots with edge-functionalization: A computational perspective }
\author{Khushboo Dange, Vaishali Roondhe, and Alok Shukla}
\affiliation{Department of Physics, Indian Institute of Technology Bombay, Powai,
Mumbai 400076, India}
\email{khushboodange@gmail.com, oshivaishali@gmail.com, shukla@iitb.ac.in}

\begin{abstract}
In this work, we have systematically investigated the structural,
electronic, vibrational and optical properties of the edge-functionalized
hg-C$_{3}$N$_{4}$ quantum dots with the aim of exploring their possible
applications in solar cells and other optoelectronic devices such
as light-emitting diodes. The functional groups considered in this
work are methyl ($-CH_{3}$), fluorine ($-F$), and oxygenated groups
such as aldehyde ($-CHO$), carboxyl ($-COOH$), ketone ($-COCH_{3}$),
and hydroxyl ($-OH$) groups. The edge-functionalization resulted
in significant tuning of the electronic, vibrational, and optical
properties. Thus, their structural fingerprints are present in both
their vibrational and optical properties, thereby allowing their detection
both in the Raman as well as optical spectroscopies. It is observed
that edge functionalization broadens the energy range of optical absorption,
leading to coverage of most of the ultraviolet and visible regions.
This implies that the edge-functionalization of hg-C$_{3}$N$_{4}$
quantum dots can be used in a variety of optoelectronic devices such
as solar cells and light emitting diodes.
\end{abstract}
\maketitle

\section{Introduction}

Two-dimensional (2D) materials have attracted a lot of attention from
the research community since the successful exfoliation of graphene
\citep{novoselov2004,geim2007}. Also, the synthesis of other 2D materials
such as silicene \citep{vogt2012}, phosphorene \citep{woomer2015},
boron nitride \citep{singla2015}, carbon nitride \citep{yu2016},
silicon carbide \citep{gao2022experimental}, etc., has opened a novel
field of research in materials with useful and promising applications
in nanotechnology \citep{bhimanapati,naikoo2022,zhuang2013}. Among
these, $\pi$-conjugated 2D materials are of particular interest because
of their unique properties which make them better candidates for multiple
applications such as sensors, energy storage, electronic devices,
etc. \citep{li2013,cheng2009,qi2013,kumar2022,mike2013,zheng2014,zhang2016}.
One such material is graphitic carbon nitride (g-C$_{3}$N$_{4}$)
with a delocalized $\pi$-conjugated structure, with weak interlayer
van der Waal interactions, and strong covalent intralayer bonds \citep{wang2016},
similar to graphene. Among all carbon nitride structures, g-C$_{3}$N$_{4}$
is the most stable allotrope as compared to its other phases such
as the cubic, semi-cubic, $\alpha$, and $\beta$ phases \citep{zhang2001}.
Furthermore, it possesses the smallest band gap as compared to its
other allotropes\citep{wang2017}. The heptazine phase has an indirect
band gap of 2.7 eV, while triazine phase possess a direct band gap
of 2.9 eV \citep{ismael2019,XU2012}. Additionally, an atomically
thin infinite 2D monolayer of g-C$_{3}$N$_{4}$ also exists \citep{yu2016},
and its band gap computed using the first-principles density functional
theory (DFT) is reported to be 2.1 eV \citep{Wang2009}. This material
has gained widespread attention because of several reasons such as
its direct band gap in the visible region, low cost, earth abundance,
easy synthesis, metal-free nature, physiochemical stability, and good
thermal stability \citep{darkwah2018}. 2D g-C$_{3}$N$_{4}$ covers
a wide range of possible applications including photocatalytic properties
both for the hydrogen evolution reaction (HER) as well as oxygen evolution
reaction (OER), bioimaging, and in photoelectronic devices\citep{darkwah2018,zheng2014,huang2017,zhang2013,dong2014}.
In spite of all these fascinating properties, pure g-C$_{3}$N$_{4}$
has its own limitations as a photocatalytic material, low electrical
conductivity, and inefficient utilization of solar energy due to its
wide band gap, thereby restricting its applicability \citep{zheng2014,huang2017,dong2014}.
Patnaik et al. \citep{patnaik2018} recently reviewed the advances
in designing Ag modified g-C$_{3}$N$_{4}$ based nanocomposites to
enhance its photocatalytic activity. Tian et al. \citep{tian2014}
synthesized a g-C$_{3}$N$_{4}$--BiVO$_{4}$ heterojunction which
delivers high photocatalytic performance. Chemical functionalization
such as carboxylation, sulfonation, amidation, and phosphorylation
as well as substitutional doping with B, C, N, S, and O resulted in
significant modifications of the electronic and optical properties
of g-C$_{3}$N$_{4}$ \citep{majdoub2020,ullah2020,chen2021}. g-C$_{3}$N$_{4}$
has been widely used in dye-sensitized solar cells (DSSC) as a photoanode
to act as blocking layer to prevent charge recombination leading to
improved efficiency ranging from 2.01--8.07 \% \citep{YANG2021}.
It has also been added in various perovskite solar cells with the
purpose of improving the coupling between the pervoskite layer and
the hole transport material leading to increased efficiency in the
range 12.85 -- 20.3 \% \citep{YANG2021}.

Nowadays, research in the field of quantum dots (QDs) has received
extra attention as confinement in all dimensions (0D materials) leads
to effective tuning of electronic, optical, physical, and chemical
properties \citep{wang2016,Li2021,Xu2022}. Therefore, quantum confinement
by constructing finite structures, i.e., quantum dots of 2D g-C$_{3}$N$_{4}$
may provide us with an effective way of tuning its material properties.
Since the g-C$_{3}$N$_{4}$ sheet is a periodic arrangement of two
different types of unit cells, i.e., s-triazine and tri-s-triazine
(also known as heptazine) \citep{ullah2020}, we can obtain two types
of quantum dots from it \citep{ghashghaee2020}. Tri-s-triazine, as
the name suggests, consists of three s-triazine rings \citep{wang2012},
and it is attractive from the point-of-view of doping, adsorption,
and tunable optoelectronic properties because extra nitrogen atoms
contribute more lone pairs as compared to s-triazine-based QDs \citep{ghashghaee2020ads}.
Ghashghaee et al. \citep{ghashghaee2020qun} compared the geometric
structure and electronic properties of heptazine-based g-C$_{3}$N$_{4}$
(hg-C$_{3}$N$_{4}$) QDs and 2D g-C$_{3}$N$_{4}$. Olademehin et
al. \citep{Olademehin2021} reported the electronic and optical properties
of triangular shaped g-C$_{3}$N$_{4}$ QDs of increasing sizes, designed
using melamine (triazine) and heptazine units. Their study also demonstrated
that the carbon and nitrogen sites would be more favorable for HER
and OER, respectively. Also, the computational approach of Ullah et
al. \citep{ullah2018} gave an in-depth explanation of the better
spatial confinement of frontier orbitals and charge transfer in CNQDs
than in GQDs, leading to their enhanced photocatalytic activity. Their
investigations were also based on the triangular QDs and reported
that tuning of the optical absorption and emission depends mostly
on the size, and not on the shape. Though pure g-C\textsubscript{}$_{3}$N$_{4}$
QDs are found to be promising for solar cell devices and photocatalytic
activity, but their wide band gap limits their energy harvesting efficiency.
Both theoretical and experimental studies have shown that the HOMO-LUMO
gap of g-C$_{3}$N$_{4}$ QDs can be tuned by doping with non-metal
atoms \citep{ghashghaee2020,wu2016,bandyopadhyay2017}. Zhai et al.
\citep{zhai2018} have performed a DFT-based computational study of
pristine g-C$_{3}$N$_{4}$ QDs with the aim of understanding the
evolution of their electronic structure and optical properties as
functions of shapes and sizes. Their studies revealed that triangular
lamellar structures are more suitable candidates for superior photophysical/optical
properties, however, they did not consider functionalization. Bandhopadhyay
et al. \citep{bandyopadhyay2017} have examined the effect of functionalization
on the heterostructures composed of g-C$_{3}$N$_{4}$ QDs stacked
with graphene QD, with a single electron acceptor (carboxyl) or electron
donor (amine and hydroxyl) group. Functionalization of g-C$_{3}$N$_{4}$
QDs with the carboxyl (-COOH) and hydroxyl (-OH) groups has already
been achieved experimentally, and tunable emission properties were
obtained \citep{zhou2013,zhou2015}. To the best of our knowledge,
no theoretical study has been reported that shows the effect of functionalization
on g-C$_{3}$N$_{4}$ QDs with different electron acceptor and electron
donor groups. Also, it is reported that the synthesis of g-C$_{3}$N$_{4}$
QDs leads to a high percentage of amine edges as well as oxygenated
groups which get introduced inevitably \citep{zhou2015}. Thus, it
is important to study the effect of such functional groups on the
structural, electronic, and optical properties of g-C$_{3}$N$_{4}$
QDs. In the present work, the smallest unit of heptazine and triangular
shaped hg-C$_{3}$N$_{4}$ QDs comprising of three to six heptazine
units are considered to investigate the effect of chemical functionalization
on them. The functional groups considered in the present work include
methyl (-CH$_{3}$), fluorine (-F), and oxygenated groups such as
aldehyde (-CHO), carboxyl (-COOH), ketone (-COCH$_{3}$) and hydroxyl
(-OH) groups. The motivation for this study comes from the work of
Yunhai and collaborators on the edge-functionalized GQDs \citep{li2015}.
Their study revealed that the functional groups containing C=O double
bond are comparatively more effective in tuning the electronic and
optical properties of GQDs.

The remainder of this article is organized as follows. In the next
section, we address our computational methodology in brief, followed
by a detailed discussion of our results in section \ref{sec:Results-and-Discussion}.
Finally, we conclude our work in section \ref{sec:Conclusion} by
summarizing the key findings.

\section{Computational Details}

All the calculations presented in this work were performed within
the framework of density functional theory (DFT) \citep{hohenberg1964,kohn1965}
as implemented in the Gaussian16 package \citep{gaussian16}. The
B3LYP hybrid functional \citep{becke1992} was employed to account
for the exchange and correlation effects, coupled with the Gaussian-type
valence triple zeta 6-311G \citep{krishnan1980} basis set containing
two polarization functions (d, p). Some of the electronic and optical
properties calculations were also performed using the HSE06 functional
for comparative study. The convergence criteria was set to $10^{-8}$
Hartree to solve the Kohn-Sham equations \citep{kohn1965} self-consistently.
The geometry optimization iterations of all the considered structures
were carried out until the gradient forces on each constituent atoms
reached a minimum value of $4.5\times10^{-4}$ Hartree/Bohr. In addition,
the RMS force, maximum displacement, and RMS displacement conditions
were set at $3.0\times10^{-4}$ Hartree/Bohr, $1.8\times10^{-3}$
Bohr, and $1.2\times10^{-3}$ Bohr, respectively. The vibrational
frequencies were also calculated to ensure the stability of the optimized
structures and for none of the considered structures any imaginary
frequencies were found. GaussView6 software \citep{denn2009} was
employed for the visualization of the optimized structures and their
frontier molecular orbitals (MOs) such as the HOMO and LUMO. The partial
and total density of states were generated using the Multiwfn software
\citep{multiwfn}. After the study of the electronic properties, optical
properties were investigated using the time-dependent density functional
theory (TD-DFT) as developed by Runge and Gross \citep{runge1984,casida2009}.
The IEFPCM model \citep{IEFPCM1981} was used for the calculation
of UV-visible absorption spectra \citep{iefpcm}. 

\section{Results and Discussion}

\label{sec:Results-and-Discussion}

\subsection{Structural and Vibrational Properties}

\subsubsection*{Optimized geometries}

\begin{figure}[H]
\begin{centering}
\includegraphics[scale=0.5]{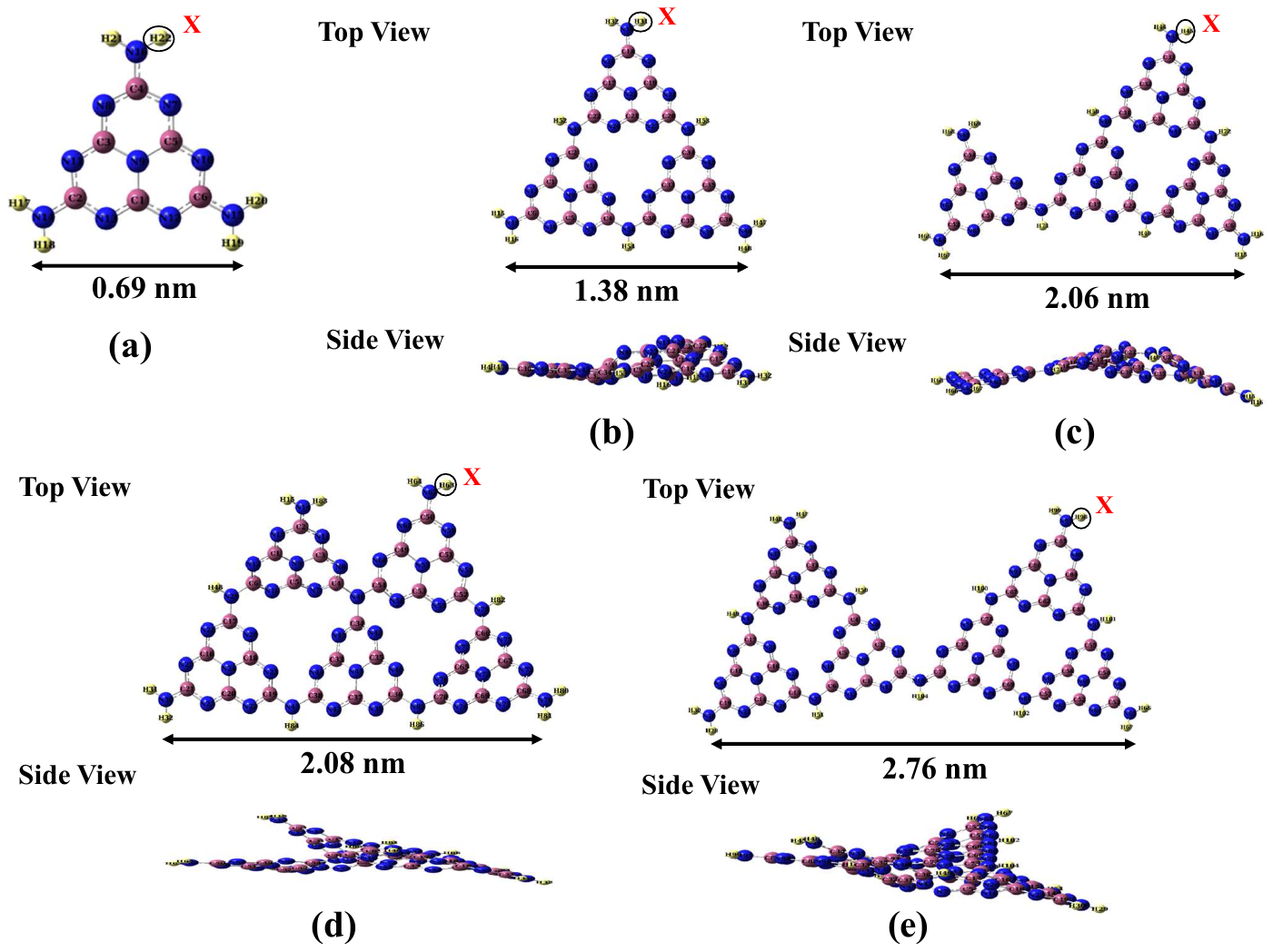}
\par\end{centering}
\caption{\label{fig:1}Optimized structures of (a) 1-X, (b) 3-X, (c) 4-X, (d)
5-X, and (e) 6-X \emph{hg}-C$_{3}$N$_{4}$quantum dots. Blue, pink,
and yellow spheres represent the nitrogen, carbon, and hydrogen atoms,
respectively, while X denotes the attached functional groups (---CH$_{3}$,
---CHO, ---COCH$_{3}$, ---COOH, ---OH, and ---F). Lateral size
of each QD is indicated by a double-arrow line.}
\end{figure}

The optimized geometries of all the considered pristine hg-C$_{3}$N$_{4}$
QDs are shown in Fig. \ref{fig:1}. The different functionalized single
heptazine units appear to remain planar (Fig. \ref{fig:1}(a)). The
calculated C--N bond lengths are in the range 1.32--1.35 Å, consistent
with the previously reported work \citep{zhai2018}. We note that
with the increase in size of the hg-C$_{3}$N$_{4}$ QDs, optimized
structures no longer remain planar and instead acquire buckled geometries.
The reason for buckling can be related to the deformation of C--N
bonds due to repulsive interactions between lone pairs present on
the nitrogen atoms \citep{zhai2018}. As the size of the hg-C$_{3}$N$_{4}$
QDs increases, the number of nitrogen atoms and the corresponding
lone pairs present in the structure increases, which results in an
increased order of buckling with the size of the hg-C$_{3}$N$_{4}$
QDs. The lateral sizes of all the optimized QDs are indicated in Fig.
\ref{fig:1}, and their values are presented in Table \ref{tab:1}.
The calculated size of the single heptazine unit is 0.69 nm, consistent
with the previously reported value \citep{zhai2018}. Interestingly,
our calculated sizes of the QDs comprising 4-6 units of heptazine
are well within the reported experimental values \citep{zhou2013,zhou2015}.
The QDs are relaxed again after attaching each of the functional groups
by replacing a hydrogen atom from an edge of their optimized structures.
The qualitative behavior of the optimized geometries of all the considered
edge-functionalized hg-C$_{3}$N$_{4}$ QDs resembles their pristine
counterpart, therefore they are shown in the same figure with one
hydrogen atom being replaced by X in Fig. \ref{fig:1}. For convenience,
the notation ``n--X'' is used to represent the studied edge-functionalized
hg-C$_{3}$N$_{4}$ QDs throughout the paper, where n denotes the
number of heptazine units present in the structure, and X represents
the attached functional groups, i.e., ---CH$_{3}$, ---CHO, ---COCH$_{3}$,
---COOH, ---OH, ---F, and Pris (pristine structure). The detailed
analysis of the optimized structures suggests that all the carbon
and nitrogen atoms present in all the pristine structures are largely
$sp^{2}$ hybridized, in spite of buckling. In the functionalized
cases, the carbon atom of oxygenated groups that contains a carbon--oxygen
double bond (---CHO, ---COCH$_{3}$, ---COOH) and is attached to
the nitrogen atom of the pristine QD moieties becomes $sp^{2}$ hybridized.
Whereas, the carbon, oxygen, and fluorine atoms of---CH$_{3}$, ---OH,
and ---F groups, respectively, become $sp^{3}$ hybridized while
forming a bond with a nitrogen atom of the pristine structures. Table
S1 of the supporting information (SI) presents the calculated bond
lengths and bond angles for atoms that are near the attached functional
groups. The changes in the C--N bond lengths for all the functionalized
hg-C$_{3}$N$_{4}$ QDs are in the range 0.02--0.05 Å, with the maximum
bond lengths for the oxygenated groups (1.38 Å) and the flourine group
(1.39 Å). The reason for such small distortions in bond length is
that only a single functional group is attached at the edge of hg-C$_{3}$N$_{4}$
QDs. Due to such small distortions, the contribution of structural
influences on the electronic and optical properties of edge-functionalized
hg-C$_{3}$N$_{4}$ QDs should be minimal.

\subsubsection*{Vibrational Properties}

In order to confirm the stability of the optimized structures, vibrational
frequency calculations are performed. Total $3n-6$ vibrational modes
are obtained for each structure, where $n$ represents the total number
of atoms present in that structure. The lack of imaginary frequencies
confirms that the structural minima are likely to have been obtained.
For all the structures considered, the minimum vibrational frequencies
are reported in Table \ref{tab:1}, which in all the cases correspond
to an out-of-plane vibrational mode. For further investigation of
the vibrational properties, we have also calculated the corresponding
Raman spectra, presented in Fig. S1 of the SI. A brief description
related to some of the unique vibrational modes is also included in
SI. 

\begin{table}[H]
\caption{\label{tab:1}Lateral size (L) and minimum vibrational frequencies
(Freq) for all the considered structures, including the pristine one
(--Pris).}

\centering{}%
\begin{tabular}{ccc|ccc}
\hline 
Structure & L (nm) & Freq (cm$^{-1}$)  & Structure & L(nm) & Freq (cm$^{-1}$) \tabularnewline
\hline 
\hline 
$1-Pris$ & 0.69 & 95 & $4-COOH$ & 2.06 & 4.01\tabularnewline
$1-CH_{3}$ & 0.69  & 55.03 & $4-OH$ & 2.06 & 6.79\tabularnewline
$1-CHO$ & 0.69 & 40.24 & $4-F$ & 2.06 & 6.81\tabularnewline
$1-COCH_{3}$ & 0.69 & 33.30 & $5-Pris$ & 2.08 & 10.09\tabularnewline
$1-COOH$ & 0.69 & 37.11 & $5-CH_{3}$ & 2.08 & 9.72\tabularnewline
$1-OH$ & 0.69 & 74.98 & $5-CHO$ & 2.08 & 10.15\tabularnewline
$1-F$ & 0.69 & 72.30 & $5-COCH_{3}$ & 2.08 & 10.08\tabularnewline
$3-Pris$ & 1.38 & 24.5 & $5-COOH$ & 2.08 & 10.10\tabularnewline
$3-CH_{3}$ & 1.38 & 22.07 & $5-OH$ & 2.08 & 9.96\tabularnewline
$3-CHO$ & 1.38 & 19.20 & $5-F$ & 2.08 & 10.14\tabularnewline
$3-COCH_{3}$ & 1.38 & 16.71 & $6-Pris$ & 2.76 & 3.62\tabularnewline
$3-COOH$ & 1.38 & 17.24 & $6-CH_{3}$ & 2.76 & 3.83\tabularnewline
$3-OH$ & 1.38 & 22.13 & $6-CHO$ & 2.76 & 3.94\tabularnewline
$3-F$ & 1.38 & 21.85 & $6-COCH_{3}$ & 2.76 & 4.14\tabularnewline
$4-Pris$ & 2.06 & 7.81 & $6-COOH$ & 2.76 & 4.07\tabularnewline
$4-CH_{3}$ & 2.06 & 6.80 & $6-OH$ & 2.76 & 3.81\tabularnewline
$4-CHO$ & 2.06 & 5.20 & $6-F$ & 2.76 & 3.87\tabularnewline
$4-COCH_{3}$ & 2.06 & 4.57 &  &  & \tabularnewline
\hline 
\end{tabular}
\end{table}

\subsection{Electronic Properties}

After confirming the stability of the edge-functionalized hg-C$_{3}$N$_{4}$
QDs, next, we investigate their electronic properties using B3LYP
functional. In Table \ref{tab:2} and Table \ref{tab:Table2_continued}
we present highest occupied molecular orbital (HOMO) energy $E_{HOMO}$,
the lowest unoccupied molecular orbital (LUMO) energy $E_{LUMO}$,
the HOMO-LUMO energy gap $(E_{g})$, and the charge transfer.

We note that $E_{g}$ for the pristine hg-C$_{3}$N$_{4}$ QDs decreases
with the increasing size from 4.99 eV to 2.83 eV, clearly due to quantum
confinement.  The modification in $E_{g}$ for the different sizes
is related to the variation of $E_{HOMO}$ and $E_{LUMO}$ values
with the increasing size of the QD. The HOMO and LUMO levels represent
the electron donor (nucleophilic) and electron acceptor (electrophilic)
properties of the system, respectively. As the sizes of the pristine
QDs increase, LUMO energy levels get lowered continuously as is clear
from the values presented in Table \ref{tab:2} and Table \ref{tab:Table2_continued},
while for $E_{HOMO}$ values, non-monotonic decrease is observed. 

After the pristine structures, next we discuss the electronic properties
of the functionalized QDs. From Table \ref{tab:2} and Table \ref{tab:Table2_continued}
it is obvious that edge-functionalization causes significant changes
in the values of HOMO/LUMO energies and consequently in the HOMO-LUMO
gaps of the QDs. It is observed that for all the edge-functionalized
hg-C$_{3}$N$_{4}$ QDs (except $6-X$ QDs), the $E_{HOMO}$ values
of those functionalized with the $-CH_{3}$ group increase with respect
to their pristine counterparts, which implies increment in their electron-donor
ability. However, functionalization with other groups leads to reduced
$E_{HOMO}$ values, indicating reduced electron-donor abilities of
the corresponding QDs. Similarly, $E_{LUMO}$ values of the QDs functionalized
with the $-CH_{3}$ are seen to increase with respect to the pristine
QDs, suggesting their reduced electron-acceptor ability. However,
in the case of other functional groups, $E_{LUMO}$ values are seen
to decrease with respect to the corresponding pristine QDs implying
increased electron-acceptor ability. For the case of $6-X$ QDs, we
observe a different behavior; $E_{HOMO}$ values get lowered and $E_{LUMO}$
values are increased for all the groups compared to their pristine
counterparts. Thus, the uneven shifting of the HOMO and LUMO levels
resulted in the tuning of $E_{g}$ which depends on the following
two factors: (a) frontier orbital interaction (FOI), and (b) charge
transfer. According to frontier molecular orbital (FMO) theory \citep{fleming1982,houk1977},
interaction between frontier orbitals (HOMO and LUMO) leads to hybridization
and reduces the energy gap between them. However, charge transfer
from the QD moiety to the attached functional group leads to a reduction
in the screening of electrons. The enhancement of electronic screening
with the increase in electron density and vice versa has already been
reported previously in the literature \citep{tetsuka2020,mak2014}.
The reduction in electronic screening will increase the electron-electron
interaction, which in turn increases the energy gap. Therefore, the
tuned $E_{g}$ depends on the competition between FOI and the charge
transfer \citep{li2015}. 

We have performed the Mulliken charge analysis and then calculated
the amount of charge transfer from the hg-C$_{3}$N$_{4}$ QD moiety
to the attached functional group or vice-versa. The calculated charge
transfers for all the pristine and functionalized structures are presented
in Table \ref{tab:2} and Table \ref{tab:Table2_continued}. We note
that for the pristine QDs, the charge transfer is between the QD moiety
and that H-atom which is replaced by a functional group in the case
of functionalized QDs. If the charge is transferred from the QD moiety
to the attached functional group or the H atom, i.e., there is an
electron transfer from the H atom or the functional group to the QD
moiety, the charge transfer is assigned a positive sign. However,
if there is a net electron transfer in the opposite direction, i.e.,
from the QD moiety to the H atom or the functional group, the charge
transfer is assigned a negative sign. We note that the charge transfer
as defined above is positive in all the cases except for the functional
groups $-OH$ and $-F$, for which it is negative. The $-OH$ and
$-F$ groups because of their high electronegativities gain electrons
from the QD moiety, which justifies their electron withdrawing nature,
also reported for the GQDs edge-functionalized with these two groups
\citep{li2015}. Also, the amount of electron transfer to the $-OH$
group is less than that to the $-F$ group because comparatively speaking
--F group is more electronegative than the $-OH$ group.  Further,
the positive value of charge transfer in the case of $-CH_{3}$ group
is in accordance with its electron donating nature, reported for the
$-CH_{3}$ functionalized GQDs also \citep{li2015}. The amount of
charge transfer reveals the extent to which $E_{g}$ increases. Considering
the case of $6-CHO$ and $6-COCH_{3}$, charge transfer is more in
case of $-CHO$ group, which leads to a large $E_{g}$. As stated
earlier, the effects of structural distortions are minimal in this
work; the increment or decrement of $E_{g}$ is induced by the FOI
and charge transfer. The resultant $E_{g}$ for $1-CH_{3}$ QD gets
increased as compared to its pristine counterpart, while it gets lowered
in the case of other $1-X$ structures. The opposite trend is noticed
in $5-X$ QDs, as $E_{g}$ is reduced only for $5-CH_{3}$ QD. In
cases of $3-X$ and $4-X$ QDs, $E_{g}$ is reduced compared to their
pristine counterparts, except for $-OH$ functionalized cases. Also,
for $3-CH_{3}$ QD, no change is noticed, which implies that the effective
contribution of charge transfer and FOI after functionalization is
equal, and thus the cancellation of their effects takes place. In
case of $6-X$ QDs, edge-functionalization resulted in increased $E_{g}$
for all the considered functional groups. The reason for larger $E_{g}$
than their pristine counterpart in some of the cases implies that
the influence of the charge transfer is greater than the effect of
FOI. However, reduced $E_{g}$ in other cases reveals that the effective
contribution of FOI and charge transfer is such that the FOI dominates.

Furthermore, we have also calculated $E_{HOMO}$, $E_{LUMO}$, and
$E_{g}$ values using the HSE06 functional (Tables \ref{tab:2} and
\ref{tab:Table2_continued}) for comparision. In this case also, uneven
shifting of $E_{HOMO}$ and $E_{LUMO}$ values is observed as depicted
in Tables \ref{tab:2} and \ref{tab:Table2_continued}. Compared to
the B3LYP results, the HSE06 functional based $E_{g}$ values are
lower for all the considered structures. The $E_{g}$ obtained for
1-Pris (4.67 eV) and 3-Pris (3.55 eV) QDs using HSE06 functional are
relatively closer to the values reported in the literature \citep{Olademehin2021}.
The trends observed in the shifting of the HOMO and LUMO levels for
the pristine and corresponding functionalized cases are similar to
those of the B3LYP based results.

\begin{table}[H]
\caption{\label{tab:2}Calculated energies of the HOMO $(E_{HOMO})$, LUMO
($E_{LUMO}$), energy gap $(E_{g})$, and the charge transfer from
the QD moiety to the attached functional group or vice-versa for $1-X$,
$3-X$, and $4-X$ structures. $Pris$ indicates the pristine QD,
without any attached functional group.}

\centering{}%
\begin{tabular}{cccccccc}
\toprule 
Structures & \multicolumn{2}{c}{$E_{HOMO}$ (eV)} & \multicolumn{2}{c}{$E_{LUMO}$ (eV)} & \multicolumn{2}{c}{$E_{g}$ (eV)} & Charge Transfer (e)\tabularnewline
\midrule 
 & B3LYP & HSE06 & B3LYP & HSE06 & B3LYP & HSE06 & B3LYP\tabularnewline
\midrule 
$1-Pris$ & -6.25 & -6.11 & -1.26 & -1.47 & 4.99 & 4.67 & 0.25\tabularnewline
$1-CH_{3}$ & -6.20 & -6.23 & -1.19 & -1.40 & 5.01 & 4.83 & 0.17\tabularnewline
$1-CHO$ & -6.56 & -6.42 & -1.91 & -2.12 & 4.65 & 4.30 & 0.02\tabularnewline
$1-COCH_{3}$ & -6.46 & -6.32 & -1.76 & -1.97 & 4.70 & 4.35 & 0.07\tabularnewline
$1-COOH$ & -6.52 & -6.39 & -1.77 & -1.98 & 4.75 & 4.40 & 0.03\tabularnewline
$1-OH$ & -6.36 & -6.37 & -1.48 & -1.48 & 4.88 & 4.89 & -0.01\tabularnewline
$1-F$ & -6.61 & -6.47 & -1.79 & -2.00 & 4.82 & 4.47 & -0.19\tabularnewline
$3-Pris$ & -6.31 & -6.19 & -2.40 & -2.64 & 3.91 & 3.55 & 0.23\tabularnewline
$3-CH_{3}$ & -6.28 & -6.15 & -2.37 & -2.60 & 3.91 & 3.55 & 0.20\tabularnewline
$3-CHO$ & -6.46 & -6.34 & -2.65 & -2.89 & 3.81 & 3.45 & 0.06\tabularnewline
$3-COCH_{3}$ & -6.41 & -6.29 & -2.57 & -2.80 & 3.84 & 3.48 & 0.09\tabularnewline
$3-COOH$ & -6.44 & -6.32 & -2.60 & -2.84 & 3.84 & 3.48 & 0.04\tabularnewline
$3-OH$ & -6.41 & -6.29 & -2.49 & -2.72 & 3.92 & 3.57 & -0.01\tabularnewline
$3-F$ & -6.49 & -6.37 & -2.64 & -2.87 & 3.85 & 3.49 & -0.16\tabularnewline
$4-Pris$ & -6.36 & -6.24 & -2.56 & -2.80 & 3.80 & 3.44 & 0.23\tabularnewline
$4-CH_{3}$ & -6.32 & -6.21 & -2.53 & -2.77 & 3.79 & 3.44 & 0.22\tabularnewline
$4-CHO$ & -6.50 & -6.38 & --2.76 & -3.00 & 3.74 & 3.38 & 0.11\tabularnewline
$4-COCH_{3}$ & -6.45 & -6.33 & -2.70 & -2.93 & 3.75 & 3.40 & 0.13\tabularnewline
$4-COOH$ & -6.48 & -6.36 & -2.72 & -2.96 & 3.76 & 3.41 & 0.09\tabularnewline
$4-OH$ & -6.45 & -6.33 & -2.63 & -2.87 & 3.82 & 3.46 & -0.01\tabularnewline
$4-F$ & -6.52 & -6.40 & -2.75 & -2.99 & 3.76 & 3.41 & -0.14\tabularnewline
\bottomrule
\end{tabular}
\end{table}

\begin{table}[H]

\caption{\label{tab:Table2_continued}Calculated energies of the HOMO $(E_{HOMO})$,
LUMO ($E_{LUMO}$), energy gap $(E_{g})$, and the charge transfer
from the QD moiety to the attached functional group or vice-versa
for $5-X$ and $6-X$ structures. $Pris$ indicates the pristine QD,
without any attached functional group.}

\centering{}%
\begin{tabular}{cccccccc}
\toprule 
Structures & \multicolumn{2}{c}{$E_{HOMO}$ (eV)} & \multicolumn{2}{c}{$E_{LUMO}$ (eV)} & \multicolumn{2}{c}{$E_{g}$ (eV)} & Charge Transfer (e)\tabularnewline
\midrule 
 & B3LYP & HSE06 & B3LYP & HSE06 & B3LYP & HSE06 & B3LYP\tabularnewline
\midrule 
$5-Pris$ & -6.28 & -6.16 & -2.76 & -3.01 & 3.52 & 3.15 & 0.28\tabularnewline
$5-CH_{3}$ & -6.25 & -6.13 & -2.75 & -2.99 & 3.50 & 3.14 & 0.21\tabularnewline
$5-CHO$ & -6.44 & -6.32 & -2.88 & -3.12 & 3.56 & 3.20 & 0.05\tabularnewline
$5-COCH_{3}$ & -6.39 & -6.27 & -2.84 & -3.09 & 3.55 & 3.18 & 0.09\tabularnewline
$5-COOH$ & -6.41 & -6.30 & -2.85 & -3.10 & 3.56 & 3.20 & 0.04\tabularnewline
$5-OH$ & -6.38 & -6.27 & -2.82 & -3.07 & 3.56 & 3.20 & -0.01\tabularnewline
$5-F$ & -6.46 & -6.34 & -2.88 & -3.13 & 3.58 & 3.22 & -0.14\tabularnewline
$6-Pris$ & -6.33 & -5.91 & -3.50 & -3.53 & 2.83 & 2.38 & 0.28\tabularnewline
$6-CH_{3}$ & -6.38 & -6.26 & -2.77 & -3.01 & 3.61 & 3.25 & 0.24\tabularnewline
$6-CHO$ & -6.45 & -6.34 & -2.90 & -3.14 & 3.55 & 3.19 & 0.11\tabularnewline
$6-COCH_{3}$ & -6.44 & -6.33 & -2.86 & -3.10 & 3.58 & 3.22 & 0.15\tabularnewline
$6-COOH$ & -6.45 & -6.33 & -2.88 & -3.02 & 3.57 & 3.21 & 0.09\tabularnewline
$6-OH$ & -6.43 & -6.32 & -2.84 & -3.08 & 3.59 & 3.23 & -0.01\tabularnewline
$6-F$ & -6.45 & -6.33 & -2.90 & -3.15 & 3.55 & 3.19 & -0.13\tabularnewline
\bottomrule
\end{tabular}
\end{table}

\begin{figure}[H]
\begin{centering}
\includegraphics[scale=0.5]{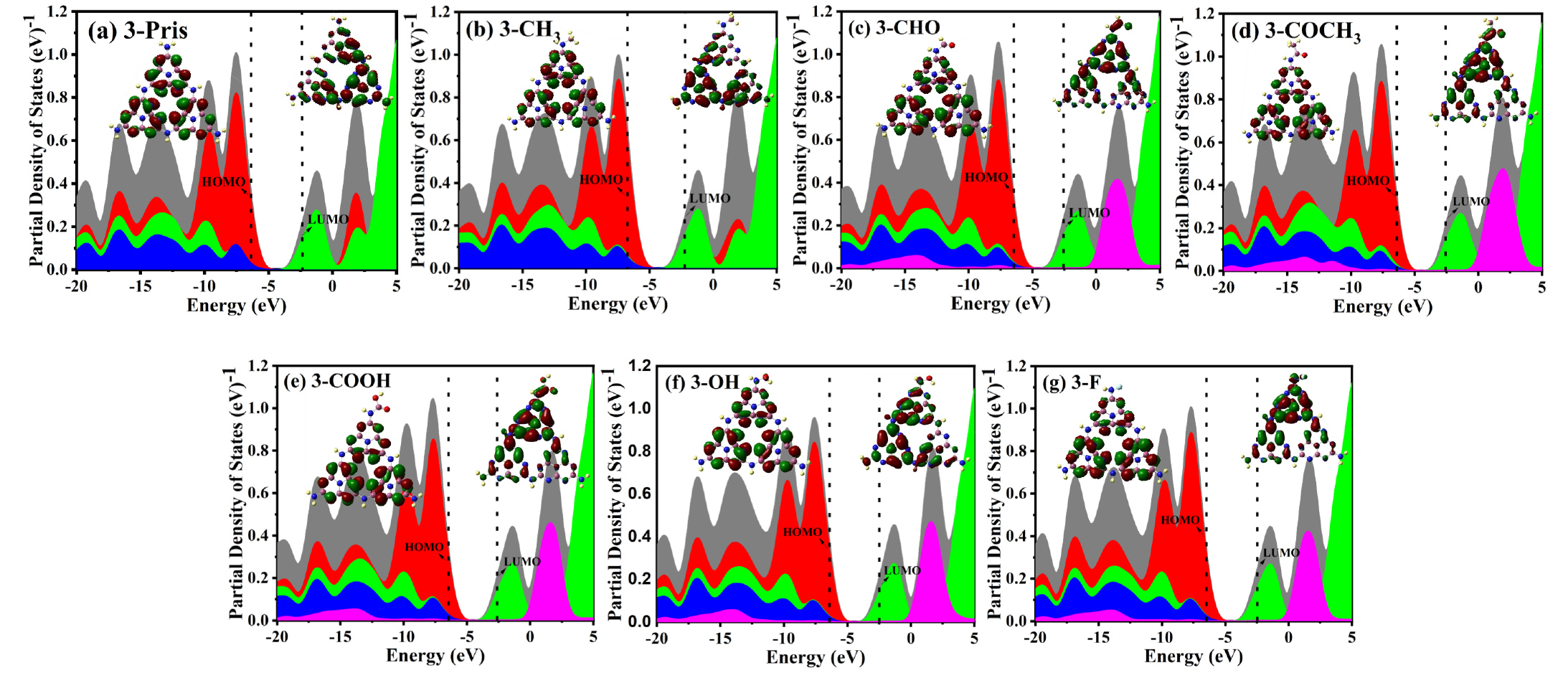}
\par\end{centering}
\caption{\label{fig:3}Total and partial density of states plots of (a) $3-Pris$,
(b) $3-CH_{3}$, (c) $3-CHO$, (d) $3-COCH_{3}$, (e) $3-COOH$, (f)
$3-OH$, and (g) $3-F$ structures. Grey color represents the TDOS.
Red, green, blue, and magenta colors represent the contributions of
H, N, C, and O or F atoms, respectively. The corresponding HOMO and
LUMO are shown in the inset of each plot.}

\end{figure}

To further investigate the electronic properties, both the total density
of states (TDOS) and partial density of states (PDOS) are calculated
for each of the edge-functionalized $hg-C_{3}N_{4}$ QDs using the
B3LYP functional, and the results are plotted in Fig. \ref{fig:3}
for the 3--X QDs, and in Figs. S2---S5 of the SI for 1--X, 4--X,
5--X, and 6--X QDs, respectively. As shown in Fig. \ref{fig:3},
the TDOS plots for $3-X$ QDs have two common features, i.e., (a)
there are five peaks visible in the occupied-orbital region, and (b)
the three peaks (the highest one is hidden in the green region) in
the unoccupied-orbital region. As is obvious from Figs. S3---S5 of
the SI, that the similar trends in TDOS are obtained for $4-X$, $5-X$,
and $6-X$ QDs also. However, for $1-X$ QDs a shoulder peak corresponding
to the third peak of the occupied region is clearly visible as depicted
in Fig. S2 of the SI. The PDOS provides the contribution of each constituent
atom individually to the TDOS. It is clear from Fig. \ref{fig:3}
that in the case of $3-X$ QDs, in the occupied region, the maximum
contribution to TDOS is from the hydrogen atoms, followed by nitrogen
and carbon atoms whereas in the unoccupied region, nitrogen atoms
contribute the most for all the QDs, while H atoms make the next most
important contributions for 3--Pris and 3--CH$_{3}$. For the QDs
functionalized with the O-based groups or F atom, O and F atoms also
contribute significantly to the TDOS in the unoccupied region for
the 3--X QDs (see Figs. \ref{fig:3}(c)---(g)). Similar behavior
of PDOS is obtained for $4-X$, $5-X$, and $6-X$ QDs, presented
in Fig. S3--S5 of the SI, respectively. In the case of $1-X$ QDs
(Fig. S2 of the SI), in addition to the above behavior, a minor contribution
of carbon atoms (hidden in the magenta region in the case of oxygenated
and fluorinated groups) is also present in the unoccupied region,
which is missing in other structures. It is also noted that both the
TDOS and PDOS plots are quite similar for the pristine and corresponding
functionalized structures, which indicates that the edge-functionalization
of $hg-C_{3}N_{4}$ QDs with a single functional group has a minimal
influence on the electronic structure of $hg-C_{3}N_{4}$ QDs. The
isosurfaces of the HOMO and LUMO corresponding to $3-X$ structures
are depicted in the insets of Figs. \ref{fig:3}, and corresponding
to 1---X, $4-X$, $5-X$, and $6-X$ structures are illustrated in
the inset of Figs. S2---S5 of the SI, respectively. For the single
heptazine unit $(1-Pris)$ structure {[}Fig. S2(a) of the SI{]}, it
is clear that the HOMO is localized on the nitrogen atoms, whereas
the LUMO is delocalized and mainly distributed over the C--N bonds
and located on the nitrogen atoms present at the boundary, in agreement
with the literature \citep{zhai2018,Olademehin2021}. After functionalization
of $1-Pris$ QD, the type of spatial distribution of both the HOMO
(localized) and LUMO (delocalized) remains unaffected. In addition
to this, LUMO also gets distributed over the atoms of the attached
functional groups except in the case of $-CH_{3}$ group. When we
examine the HOMO and LUMO plots of the larger QDs, we find that the
spatial distribution of both the HOMO and LUMO shows behavior similar
to that of the $1-X$ structures. We have also plotted the isosurfaces
of the HOMO and LUMO levels using the HSE06 functional for the 1-X
structures and shown in Fig. S6 of the SI. The spatial distribution
of these orbitals is similar to that obtained using the B3LYP functional.

\subsection{Optical Absorption Spectra}

\begin{figure}[H]
\begin{centering}
\includegraphics[scale=0.5]{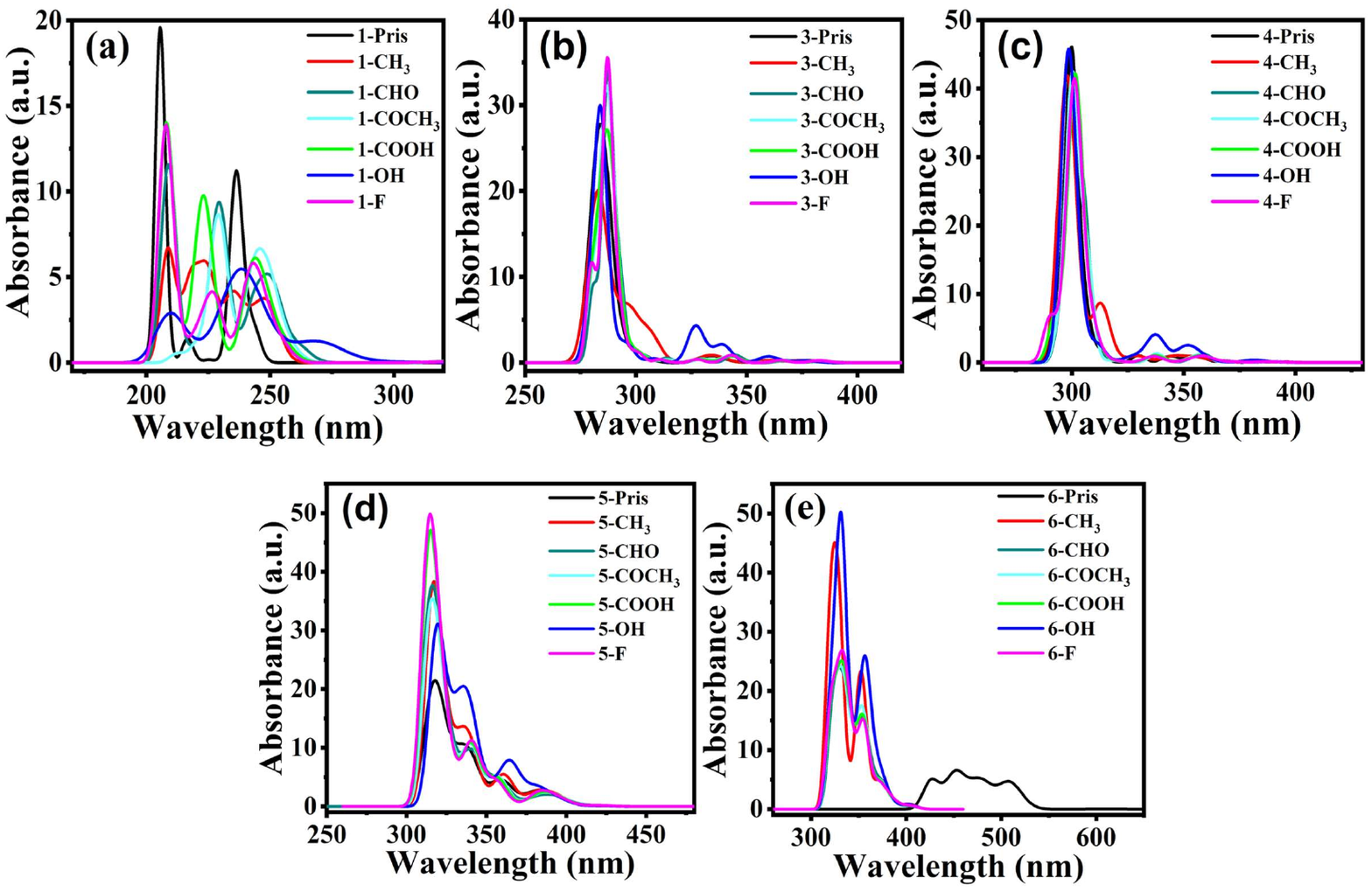}
\par\end{centering}
\caption{\label{fig:4}TD-DFT level UV-visible absorption spectra of (a) $1-X$,
(b) $3-X$, (c) $4-X$, (d) $5-X$, and (e) $6-X$ structures computed
using B3LYP, with water as the solvent.}
\end{figure}

In this section we present and discuss the optical absorption spectra
of the edge-functionalized $hg-C_{3}N_{4}$ QDs computed using the
TD-DFT approach \citep{runge1984,casida2009} under different conditions,
and compare them to those of the corresponding pristine QDs. The absorption
spectra are computed for 20 excited states. First, we calculated the
absorption spectra using B3LYP with water as the solvent (B3LYP+water),
as these QDs are found to be soluble in water \citep{wang2014}. The
calculated spectra of all the pristine and functionalized structures
using B3LYP+water are plotted in Fig. \ref{fig:4}, while their computed
optical gaps are presented in Table \ref{tab:3}. In addition, the
calculated UV-vis absorption spectra of only the pristine $hg-C_{3}N_{4}$
QDs are separately presented in Fig. S7 of the SI. 

First we note that our spectra of the pristine QDs are in good agreement
with the previously reported calculations of Zhai et al. \citep{zhai2018}.
In our calculations, the strongest absorption peaks in case of $1-Pris$
and $3-Pris$ structures are at 205 nm (6.03 eV) and 284 nm (4.36
eV), respectively, in agreement with their results \citep{zhai2018}.
As far as the size dependence of the spectra of the pristine QDs is
concerned, we note that the absorption energy range gets extended
with the increase in the size of the $hg-C_{3}N_{4}$ QDs. With the
increase in the size of the QD from 1-Pris to 6-Pris, the optical
gap ($E_{g}^{op}$) gets reduced from 5.10 eV to 2.44 eV, leading
to significant red shift also in the corresponding absorption energy
ranges. As a result, the most prominent or strongest absorption peak
red shifts from 205 nm (6.03 eV) ($1-Pris$) to 455 nm (2.72 eV) ($6-Pris$).
Thus, taking into account the calculated absorption spectra of all
the pristine QDs (1-Pris --- 6-Pris), their combined absorption range
(200--550 nm or 2.25--6.20 eV) covers most of the UV-Vis region
of the spectrum, and also lies within the range measured experimentally
\citep{wang2014}. 

From the UV-vis absorption spectra of the edge-functionalized $hg-C_{3}N_{4}$
QDs along with their pristine counterparts (see Fig. \ref{fig:4}),
it is evident that the chemical functionalization at an edge of a
given QD alters both the location as well intensity of the most intense
peak. Consequently, some of the functionalized structures undergo
a red shift, while some others experience a blue shift in the location
of the most intense peak compared to their pristine counterpart. In
addition to this, a slight variation in the total number of peaks
is also observed clearly due to the emergence of new energy levels
due to functionalization. However, the qualitative behavior of all
the edge-functionalized $hg-C_{3}N_{4}$ QDs resembles quite well
that of their pristine counterparts. In case of $-COOH$ and $-OH$
groups, our calculated absorption ranges lie within the ranges reported
experimentally \citep{zhou2013,zhou2015}. As mentioned above, the
functionalization of all the considered $hg-C_{3}N_{4}$ QDs extended
the absorption range covered as compared to the pristine structures.
For example, the absorption range of the $1-Pris$ structure 200--250
nm gets extended to 200--300 nm when functionalized with the $-OH$
group (Fig. \ref{fig:4}(a)). Thus, by functionalizing the $hg-C_{3}N_{4}$
QDs in a controlled manner we can tune their absorption ranges to
make them suitable for effective utilization of solar energy. Thus
appropriately functionalized hg-C$_{3}$N$_{4}$ QDs can be useful
in solar cells. 

\begin{figure}[H]
\begin{centering}
\includegraphics[scale=0.5]{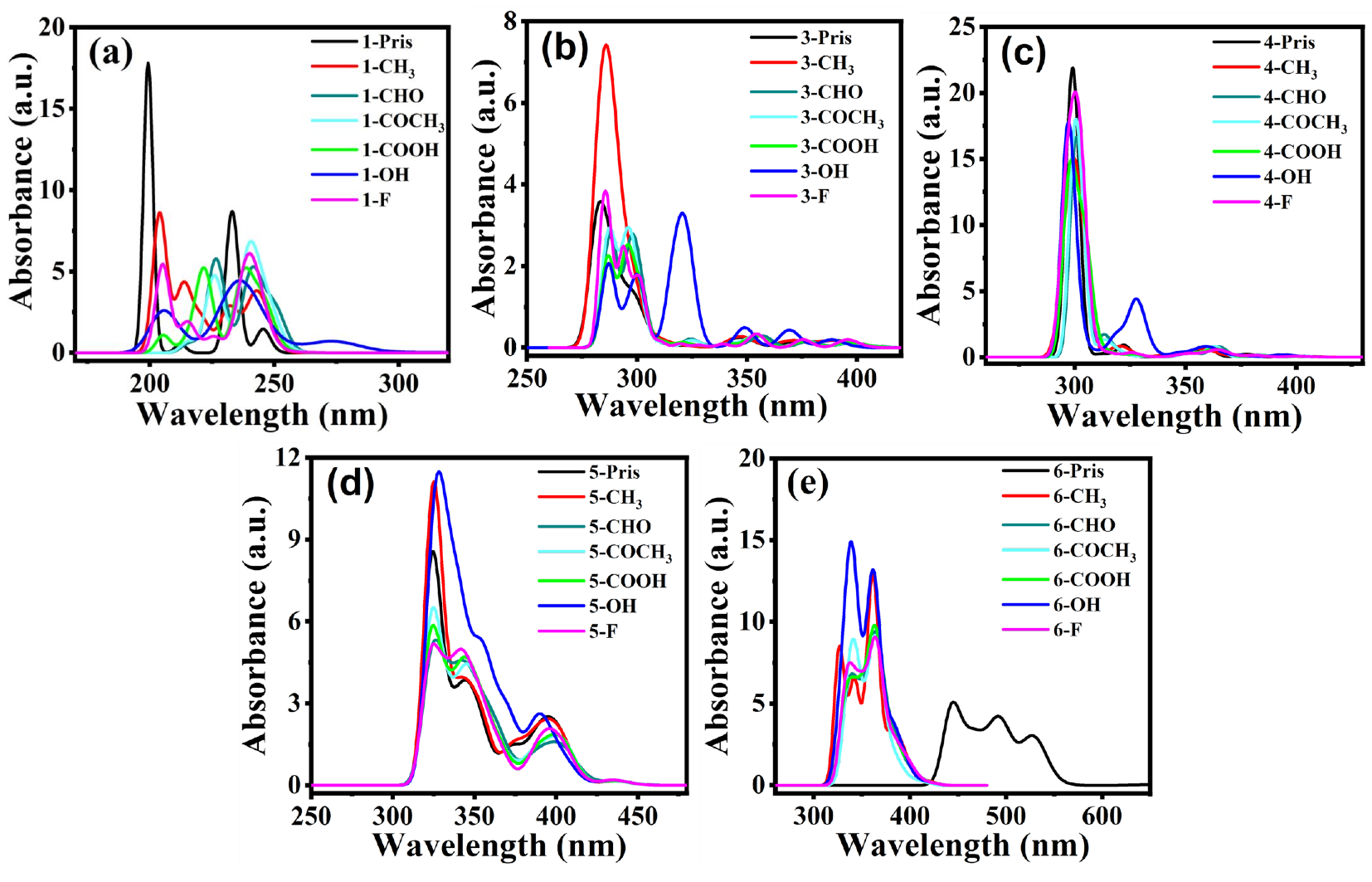}
\par\end{centering}
\caption{\textcolor{blue}{\label{fig:b3lyp_vaccum}}TD-DFT level UV-visible
absorption spectra of (a) $1-X$, (b) $3-X$, (c) $4-X$, (d) $5-X$,
and (e) $6-X$ structures computed using B3LYP+vacuum.}
\end{figure}

The optical absorption spectra are again simulated using the B3LYP
functional, but this time without including any solvent (B3LYP+vacuum),
i.e., in the gas phase. The resultant plots are illustrated in Fig.
\ref{fig:b3lyp_vaccum}, from which it is observed that the qualitative
nature of the overall spectra along with the peak positions are quite
similar to that obtained using B3LYP+water condition (Fig. \ref{fig:4}).
However, compared to the B3LYP+water-based spectra, a slight reduction
in the absorbance is observed for the 1-X QDs, while it is reduced
significantly for other structures. Interestingly, we found that the
most intense peak position corresponding to 3-Pris QD (281.64 nm)
using B3LYP+vacuum is relatively close to that reported in literature
\citep{zhai2018}. In addition, the maximum absorbance peak position
for 5-Pris QD (325.42 nm) also matches quite well with the literature
\citep{zhai2018}. We have also calculated the $E_{g}^{op}$ corresponding
to each of the edge-functionalized structures, as listed in Table
\ref{tab:3}. In the case of B3LYP+vacuum, $E_{g}^{op}$ values range
from 2.34 eV -- 5.05 eV, and compared to the earlier results (B3LYP+water),
some changes are observed. However, the total absorption range covered
by all the QDs (1-X to 6-X) in vacuum is consistent with that obtained
using water as the solvent (200 nm -- 550 nm). 

\begin{table}[H]
\caption{\label{tab:3} TD-DFT level optical gaps ($E_{g}^{op}$) of the considered
pristine as well as functionalized \emph{hg}-C$_{3}$N$_{4}$ QDs
calculated using the B3LYP functional, assuming the QDs to be in vacuum,
or dissolved in water. Additionally, the results obtained using the
HSE06 functional, and a dielectric constant 7.5 corresponding to the
g-C$_{3}$N$_{4}$ crystalline phase, are also presented.}

\centering{}%
\begin{tabular}{c|ccc||c|ccc}
\hline 
\multirow{3}{*}{Molecule} & \multicolumn{2}{c}{$E_{g}^{op}$ (eV)} &  & \multirow{3}{*}{Molecule} & \multicolumn{3}{c}{$E_{g}^{op}$(eV)}\tabularnewline
\cline{2-4} \cline{3-4} \cline{4-4} \cline{6-8} \cline{7-8} \cline{8-8} 
 & \multicolumn{2}{c|}{B3LYP} & HSE06 &  & \multicolumn{2}{c|}{B3LYP} & HSE06\tabularnewline
\cline{2-4} \cline{3-4} \cline{4-4} \cline{6-8} \cline{7-8} \cline{8-8} 
 & \multicolumn{1}{c|}{Water} & \multicolumn{1}{c|}{Vacuum} & Crystal &  & \multicolumn{1}{c|}{Water} & \multicolumn{1}{c|}{Vacuum} & Crystal\tabularnewline
\hline 
$1-Pris$ & 5.10 & 5.05 & 5.20 & $4-COOH$ & 3.45 & 3.40 & 3.45\tabularnewline
$1-CH_{3}$ & 4.96 & 4.99 & 5.06 & $4-OH$ & 3.48 & 3.42 & 3.48\tabularnewline
$1-CHO$ & 4.74 & 4.96 & 4.86 & $4-F$ & 3.45 & 3.39 & 3.45\tabularnewline
$1-COCH_{3}$ & 4.80 & 4.86 & 4.93 & $5-Pris$ & 3.45 & 3.33 & 3.44\tabularnewline
$1-COOH$ & 4.84 & 4.89 & 4.96 & $5-CH_{3}$ & 3.44 & 3.32 & 3.43\tabularnewline
$1-OH$ & 4.60 & 4.52 & 4.68 & $5-CHO$ & 3.46 & 3.42 & 3.46\tabularnewline
$1-F$ & 4.86 & 4.93 & 4.95 & $5-COCH_{3}$ & 3.47 & 3.42 & 3.46\tabularnewline
$3-Pris$ & 3.30 & 3.33 & 3.73 & $5-COOH$ & 3.47 & 3.51 & 3.47\tabularnewline
$3-CH_{3}$ & 3.44 & 3.56 & 3.42 & $5-OH$ & 3.40 & 3.51 & 3.39\tabularnewline
$3-CHO$ & 3.60 & 3.47 & 3.61 & $5-F$ & 3.47 & 3.51 & 3.47\tabularnewline
$3-COCH_{3}$ & 3.62 & 3.49 & 3.64 & $6-Pris$ & 2.44 & 2.34 & 2.47\tabularnewline
$3-COOH$ & 3.61 & 3.49 & 3.63 & $6-CH_{3}$ & 3.34 & 3.26 & 2.47\tabularnewline
$3-OH$ & 3.35 & 3.36 & 3.50 & $6-CHO$ & 3.34 & 3.25 & 3.36\tabularnewline
$3-F$ & 3.60 & 3.50 & 3.63 & $6-COCH_{3}$ & 3.34 & 3.37 & 3.43\tabularnewline
$4-Pris$ & 3.48 & 3.44 & 3.49 & $6-COOH$ & 3.34 & 3.25 & 3.44\tabularnewline
$4-CH_{3}$ & 3.48 & 3.45 & 3.51 & $6-OH$ & 3.34 & 3.25 & 3.37\tabularnewline
$4-CHO$ & 3.45 & 3.39 & 3.44 & $6-F$ & 3.34 & 3.25 & 3.43\tabularnewline
$4-COCH_{3}$ & 3.46 & 3.41 & 3.46 &  &  &  & \tabularnewline
\hline 
\end{tabular}
\end{table}

In Tables S2---S6 of the SI, we provide information related to the
excited states of the considered QDs which contribute to the most
intense peaks in the absorption spectra obtained using B3LYP+water.
This information includes the peak locations (excitation energies),
oscillator strengths ($f$), and the excited state wave functions.
In addition, Table S7 of the SI contains the same information for
the first excited state of each structure. In the TD-DFT approach,
every excited state wave function is a linear combination of several
configurations each of which corresponds to a single excitation from
an occupied orbital to a virtual one. The TD-DFT wave function of
the first excited state of each QD (pristine or functionalized) is
dominated by the configuration corresponding to the excitation of
an electron from HOMO to LUMO, denoted as $H\rightarrow L$. However,
$f$ corresponding to the first excited state for all the pristine
QDs is negligible, and even after functionalization exhibits no significant
increase, as shown in Table S7 of the SI. Therefore, the optical gap
(Table \ref{tab:3}) is larger than the excitation energies of the
first excited states for all the QDs considered in this work. 

Next we consider the optical excitations of the $4-X$ QDs, presented
in Table S4 of the SI. It is observed that the functional groups containing
the $C=O$ bond and the $-F$ group resulted in the red shift of the
absorption peaks, whereas the $-CH_{3}$ and $-OH$ groups resulted
in the blue shift compared to the pristine QD. In agreement with our
calculations, the red shift of the peaks is observed in the experiment
performed on the QDs functionalized with the $-COOH$ group \citep{zhou2013}.
The excited state leading to the most intense peak of $4-Pris$ QD
is written as $16^{1}A$, which signifies the $16^{th}$ TD-DFT singlet
excited state. As is obvious from the table, the wave function of
this excited state derives important contributions from the configurations
$H-6\rightarrow L+1$, $H-6\rightarrow L$, and $H-7\rightarrow L$.
The excited state corresponding to most intense peak in case of $4-CH_{3}$
QD is $18^{1}A$, with the wave function composed predominantly of
configurations $H-7\rightarrow L$, $H-8\rightarrow L$, and $H-5\rightarrow L$.
In a similar manner, the optical transitions corresponding to the
most intense peak are presented for each of the considered pristine
and edge-functionalized $hg-C_{3}N_{4}$ QD in those tables. In case
of $6-X$ QDs (Table S6), edge-functionalization with all the considered
groups resulted not only in an abrupt blue-shift of the most intense
peak position, but also in the range of the absorption spectrum from
the visible to the UV region after functionalization. In addition
to analyzing the orbitals involved in the transitions corresponding
to various absorption peaks, we have also studied the spatial distribution
of electrons and holes in the first excited state, and the one giving
rise to the most intense peak (the excited state with the maximum
$f$ value). For the purpose, we have used the approach of Liu et
al. \citep{e-h-den-LIU2020461} in which the hole and electron spatial
distributions are defined in terms of corresponding densities

\begin{equation}
\rho^{hole}({\bf r})=\underset{i\rightarrow a}{\sum}(w_{i}^{a})^{2}\phi_{i}({\bf r})\phi_{i}({\bf r})+\underset{i\rightarrow a}{\sum}\underset{j\neq i\rightarrow a}{\sum}w_{i}^{a}w_{j}^{a}\phi_{i}({\bf r})\phi_{j}({\bf r})
\end{equation}

\begin{equation}
\rho^{electron}({\bf r})=\underset{i\rightarrow a}{\sum}(w_{i}^{a})^{2}\phi_{a}({\bf r})\phi_{a}({\bf r})+\underset{i\rightarrow a}{\sum}\underset{i\rightarrow b\neq a}{\sum}w_{i}^{a}w_{i}^{b}\phi_{a}({\bf r})\phi_{b}({\bf r}),
\end{equation}

where $i,j$ and $a,b$ are the indices denoting to the occupied (hole)
and virtual (electron) MOs, respectively. The numbers $w_{i}^{a}$
are obtained from the TD-DFT calculations, and represent the coefficients
of the singly-excited configurations in which an electron is promoted
from the occupied MO $\phi_{i}(\vec{r})$ to the virtual MO $\phi_{a}(\vec{r})$.
We calculated the densities $\rho^{hole}(\vec{r})/\rho^{electron}(\vec{r})$
using the Multiwfn software \citep{multiwfn}, and the results are
presented in Figs. S8 -- Fig. S12 of the SI for all the considered
edge-functionalized $hg-C_{3}N_{4}$ QDs. 

\begin{figure}[H]
\begin{centering}
\includegraphics[scale=0.5]{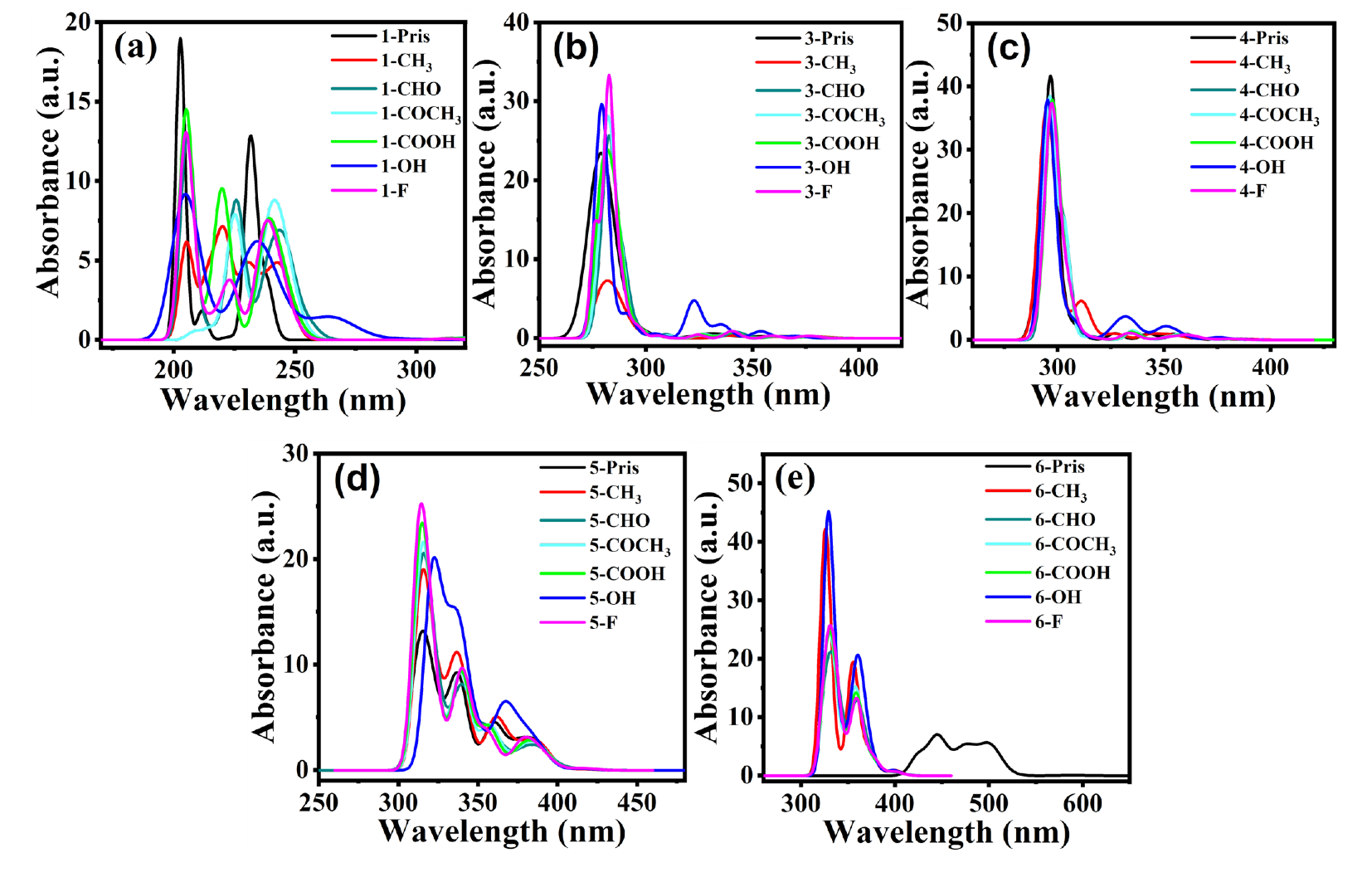}
\par\end{centering}
\caption{\textcolor{blue}{\label{UV_HSE06}}TD-DFT level UV-visible absorption
spectra of (a) $1-X$, (b) $3-X$, (c) $4-X$, (d) $5-X$, and (e)
$6-X$ structures computed using the HSE06 functional and a crystalline
environment.}

\end{figure}

After the study of optical properties using the B3LYP functional,
the UV-vis optical absorption spectra were also computed using a range
separated hybrid functional HSE06. These calculations were performed
to explore the influence of: (a) a modern range-separated hybrid functional,
and (b) crystalline environment. The crystalline form is considered
because the $g-C_{3}N_{4}$ QDs also exist in the crystalline structures
\citep{Li2018}. Therefore, during the simulation of optical absorption
spectra using the HSE06 functional, a dielectric medium relevant to
crystalline $hg-C_{3}N_{4}$ QDs is taken into consideration \citep{Zheng2017,Sun2016}.
For simulating the dielectric medium within the IEFPCM model, dielectric
constant is taken to be 7.5 which is the value corresponding to the
crystalline carbon nitrides \citep{Patra2021}. The UV-vis optical
absorption spectra computed in the crystalline environment using HSE06
functional are presented in Fig. \ref{UV_HSE06}. It is observed that
the qualitative behavior of these spectra is similar to those simulated
using B3LYP+water (Fig. \ref{fig:4}). Except for the 5-X QDs, slight
variations in the maximum absorbance values are observed. In the case
of 1-X structures, chemical functionalization resulted in the red
shifting of peaks corresponding to maximum absorbance compared to
their pristine counterpart. However, opposite trend, i.e., blue shift
is noticed after functionalization of the 6-Pris QD. Chemical functionalization
of 5-Pris QD also leads to blue shift of the peaks except $5-OH$
structure for which red shift is observed. In the case of 3-X and
4-X QDs, except $-CH_{3}$, all other functional groups lead to a
red shift of the most intense peak. Furthermore, again the complete
absorption range covered (200 nm -- 550 nm) is same as that obtained
using the B3LYP functional both with water and vacuum. Next, the $E_{g}^{op}$
values are also calculated for the HSE06+crystal condition, and the
values are reported in Table \ref{tab:3}. It is to be noticed that
for most of the structures $(E_{g}^{op})_{HSE06+crystal}>(E_{g}^{op})_{B3LYP+water}$,
while the opposite results are obtained for some of the structures. 

After the calculation of optical properties using three different
conditions, we tried to compared our obtained $E_{g}^{op}$ values
with some experimental results. The two different studies reported
that the peaks in the photoluminescence spectra of $hg-C_{3}N_{4}$
QDs lie at 367 nm (3.39 eV)\citep{zhou2015} and 467 nm (2.65 eV)
\citep{zhou2013}. The size distribution of their $hg-C_{3}N_{4}$
QDs is within the range of 2 nm - 4 nm. Compared to these experimental
findings, the optical gaps obtained in our work are 2.34 eV, 2.44
eV, and 2.47 eV using B3LYP+water, B3LYP+vacuum, and HSE06+crystal
parameters, respectively, for the biggest size (2.76 nm) pristine
quantum dot (6-Pris). 

In brief, the triangular shaped $hg-C_{3}N_{4}$ QD structures are
designed using a bottom-up approach. Starting from a single heptazine
unit, we combined more such units at the edges (by replacing one of
the edge hydrogen atoms) in a way to form larger triangular structures.
Our calculations of the optical properties of the considered triangular
shaped \emph{hg}-C\textsubscript{3}N\textsubscript{4} QDs suggest
them to be size dependent. The UV-vis absorption spectra get red shifted
with the increase in the size of the QDs due to the decrease in the
optical gap. Chemical functionalization of the hg-C3N4 QDs resulted
in the shifts of the most intense peaks in the absorption spectra
compared to their pristine counterpart. Some of the functionalized
structures undergo a red shift, while some others experience a blue
shift. In addition, the functionalization of all the considered \emph{hg}-C\textsubscript{3}N\textsubscript{4}
QDs extended the absorption range covered by their corresponding pristine
structures. Therefore, edge-functionalization is an effective approach
to enhance the photophysical properties of \emph{hg}-C\textsubscript{3}N\textsubscript{4}
QDs by tuning their absorption ranges to make them more suitable for
utilization of the solar energy. The significant reduction in band
gap with the increasing size and a better optical response also make
them potential candidate for photocatalytic applications.

\section{Conclusion}

\label{sec:Conclusion}

To summarize, in this work we have presented an exhaustive first-principles
DFT-based study of the electronic, vibrational, and optical properties
of pristine and functionalized quantum dots of a novel 2D material
g-C$_{3}$N$_{4}$. Triangular structures of increasing sizes derived
from heptazine were considered, and their geometries were optimized,
followed by a check of their dynamic stability by performing a detailed
vibrational frequency analysis. Additionally, the Raman spectrum of
each considered structure was computed and analyzed. Electronic properties,
such as the HOMO-LUMO energy gap, and the charge transfer were also
studied, and it was observed that the edge-functionalization is an
effective way of tuning the electronic properties of the $hg-C_{3}N_{4}$
QDs. Further, the influence of functionalization was also studied
by comparing both the partial and total density of states of the functionalized
and pristine structures. 

Using the TD-DFT methodology, the UV-vis absorption spectra of all
the structures were computed and analyzed in detail using two different
hybrid functionals with different conditions. We found that most of
the ultraviolet region gets covered in the pristine structures itself,
which gets shifted to the visible region in the case of $6-Pris$
QD due to the increase in size. Edge functionalization further extended
the absorption range covered by the corresponding pristine structures
suggesting that the edge-functionalized $hg-C_{3}N_{4}$ QDs will
operate in a wide energy range and will be effective in enhancing
the efficiency of solar cells. Moreover, their excellent optical properties
make them a potential candidate for other optoelectronic devices such
as light-emitting diodes operating both in the visible and ultraviolet
range. Our results for the UV-vis spectra obtained in the case of
carboxylic and hydroxyl groups are consistent with those obtained
experimentally. Consequently, we hope that this theoretical study
of the absorption spectra of edge functionalized $hg-C_{3}N_{4}$
QDs using other functional groups will guide future experimental endeavors.
Also our idea of using two different functionals and three different
environmental parameters (vacuum, water, and crystalline environment)
to explore their optical properties will be helpful for various experimental
studies. 

\section*{Conflicts of Interest}

There are no conflicts of interest to declare.
\begin{acknowledgments}
One of the authors, K.D. acknowledges financial assistance from Prime
Minister Research Fellowship (PMRF award ID-1302054). V.R. acknowledges
the support through the Institute Post-Doctoral Fellowship (IPDF)
of Indian Institute of Technology Bombay. 
\end{acknowledgments}

\bibliographystyle{ieeetr}
\bibliography{Reference_C3N4}

\newpage\includegraphics[scale=0.85]{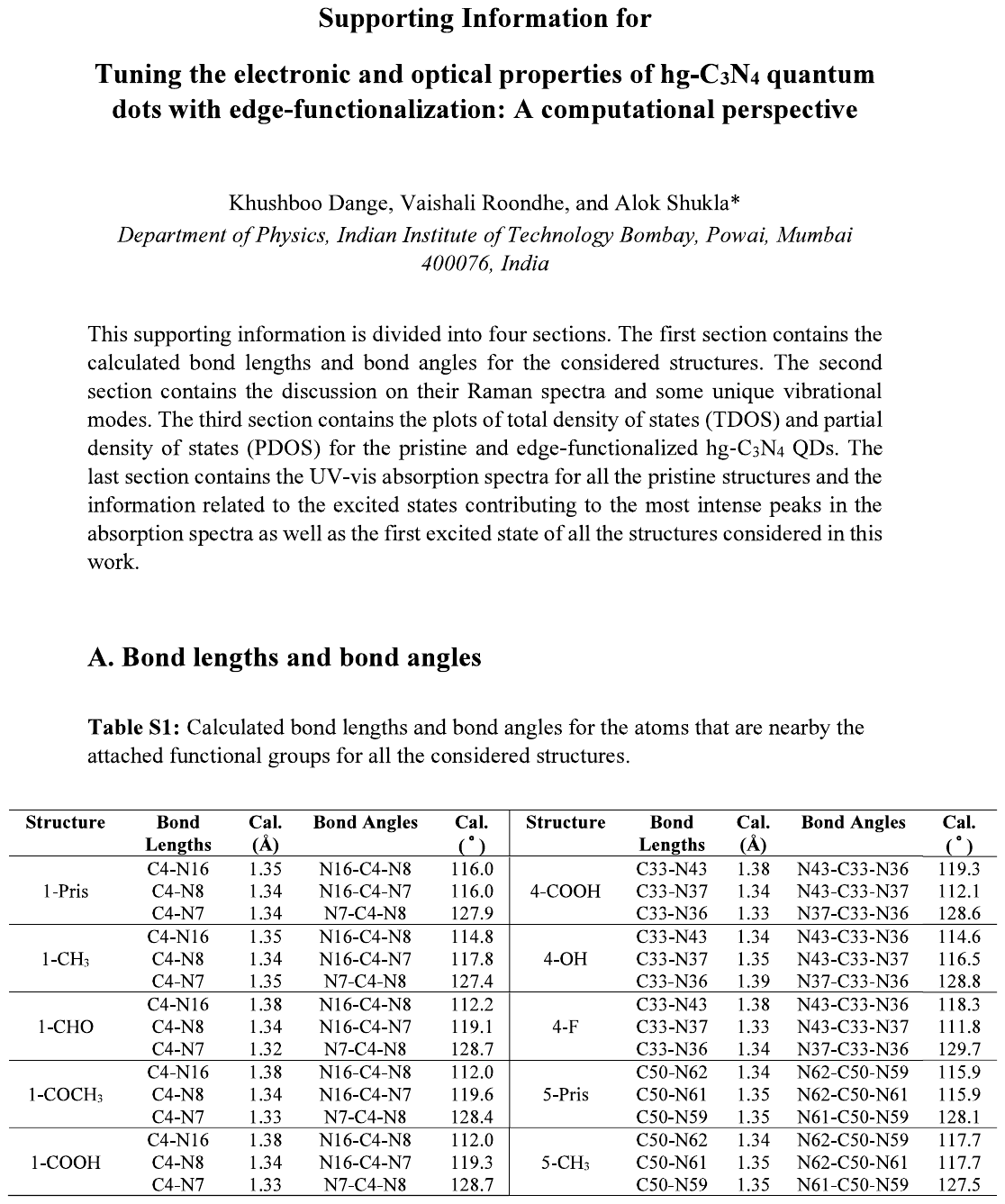}

\includegraphics[scale=0.85]{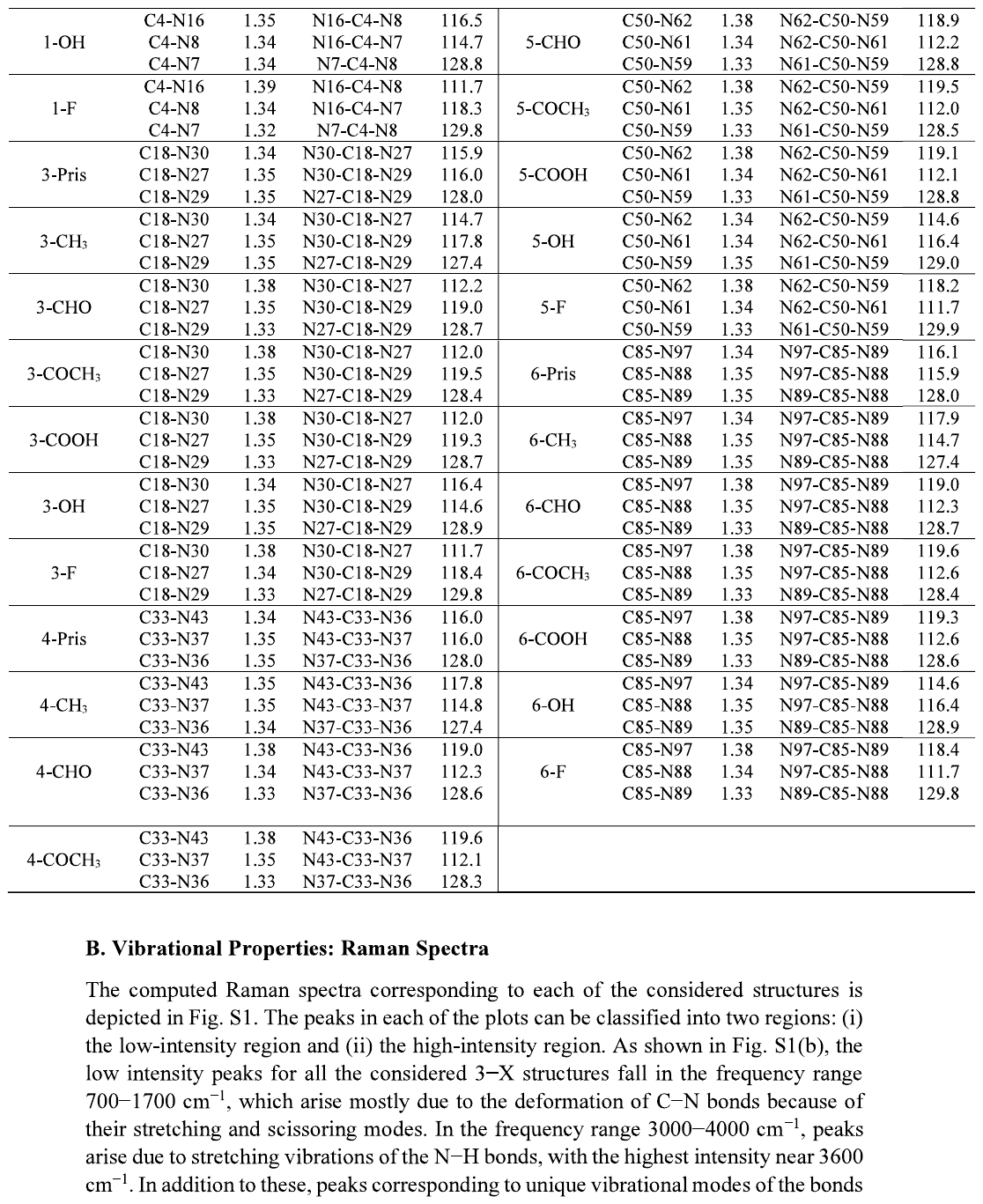}

\includegraphics[scale=0.85]{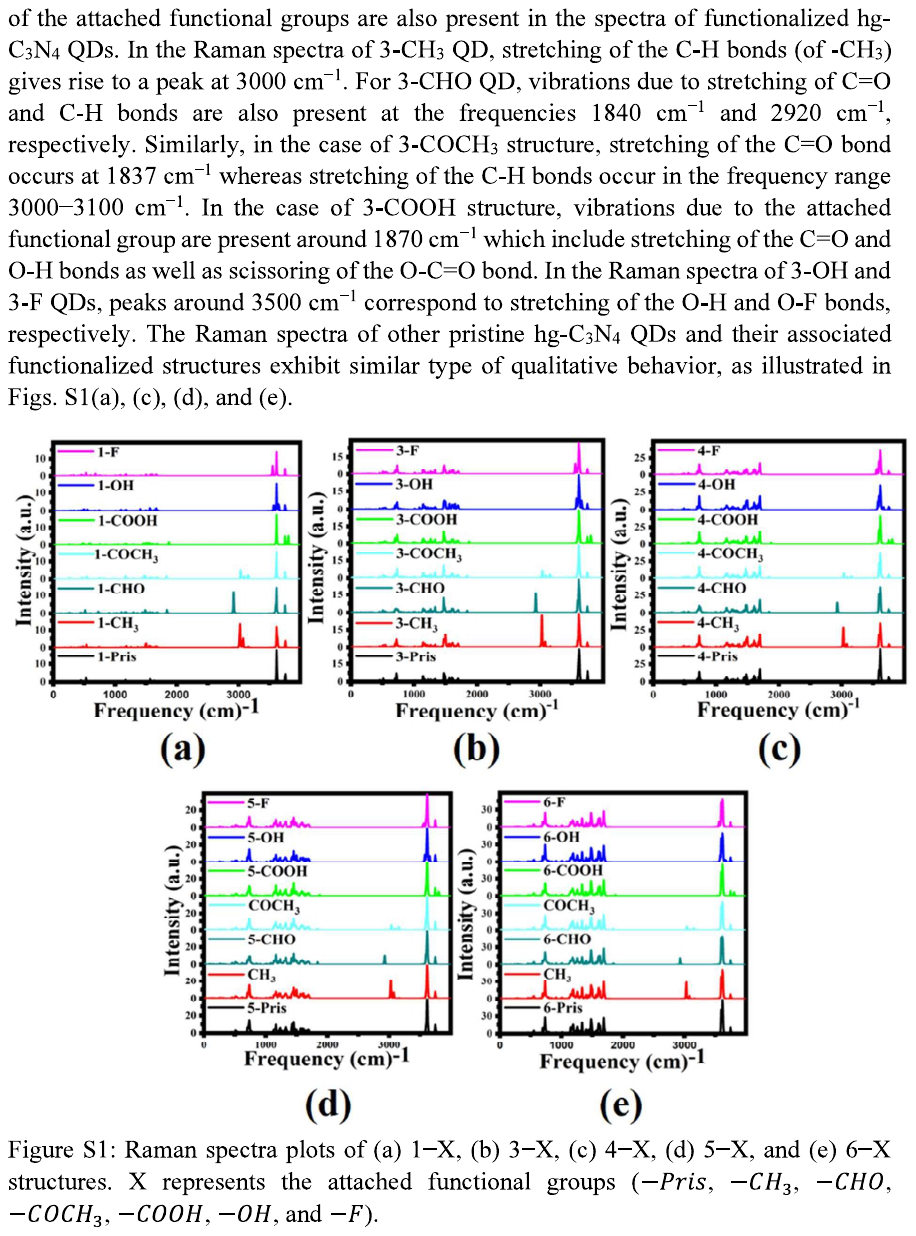}

\includegraphics[scale=0.85]{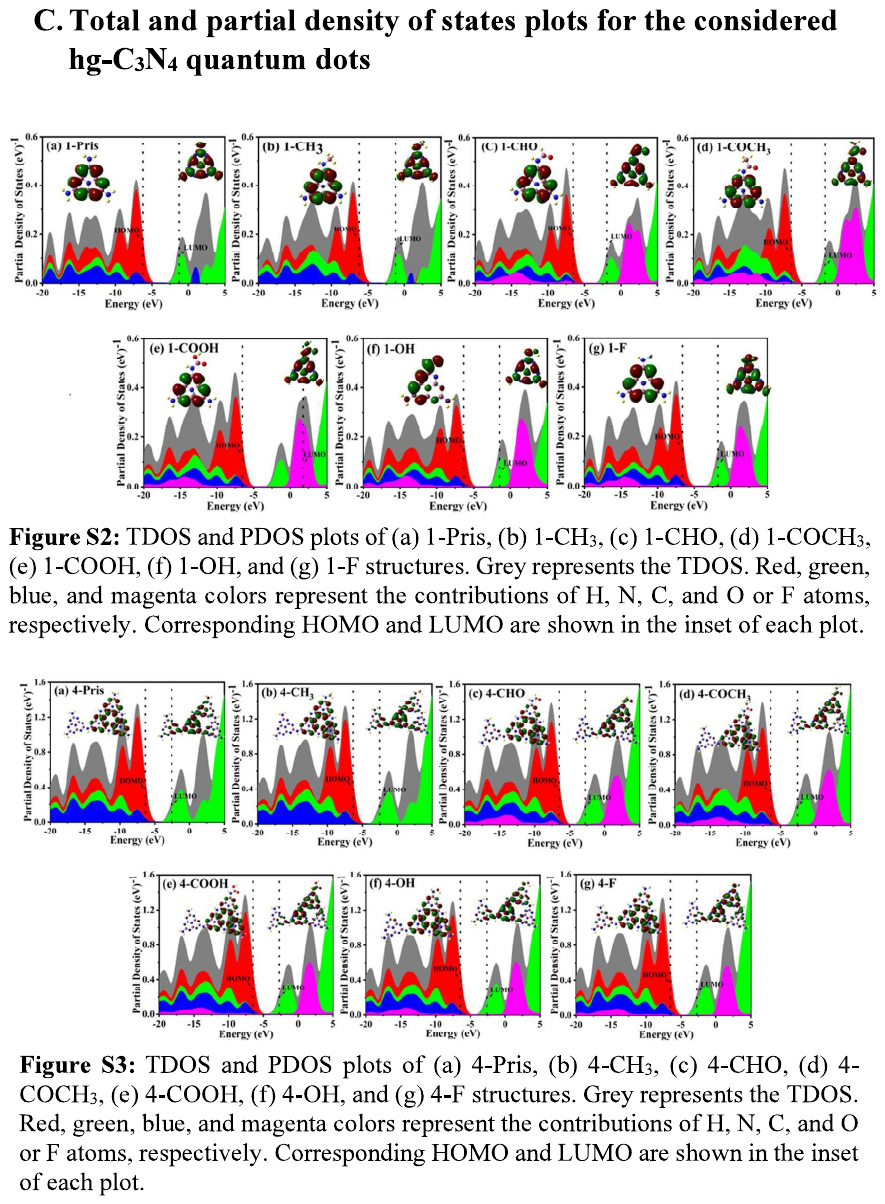}

\includegraphics[scale=0.85]{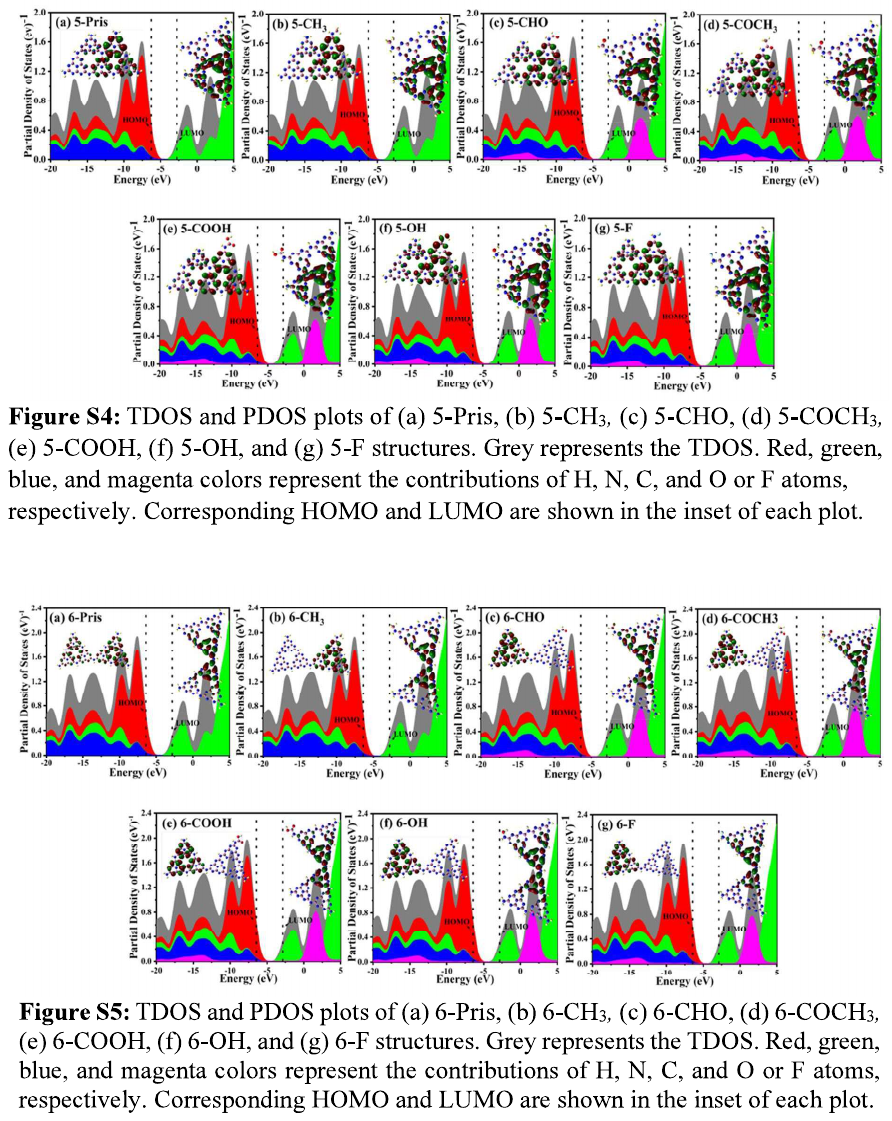}

\includegraphics[scale=0.85]{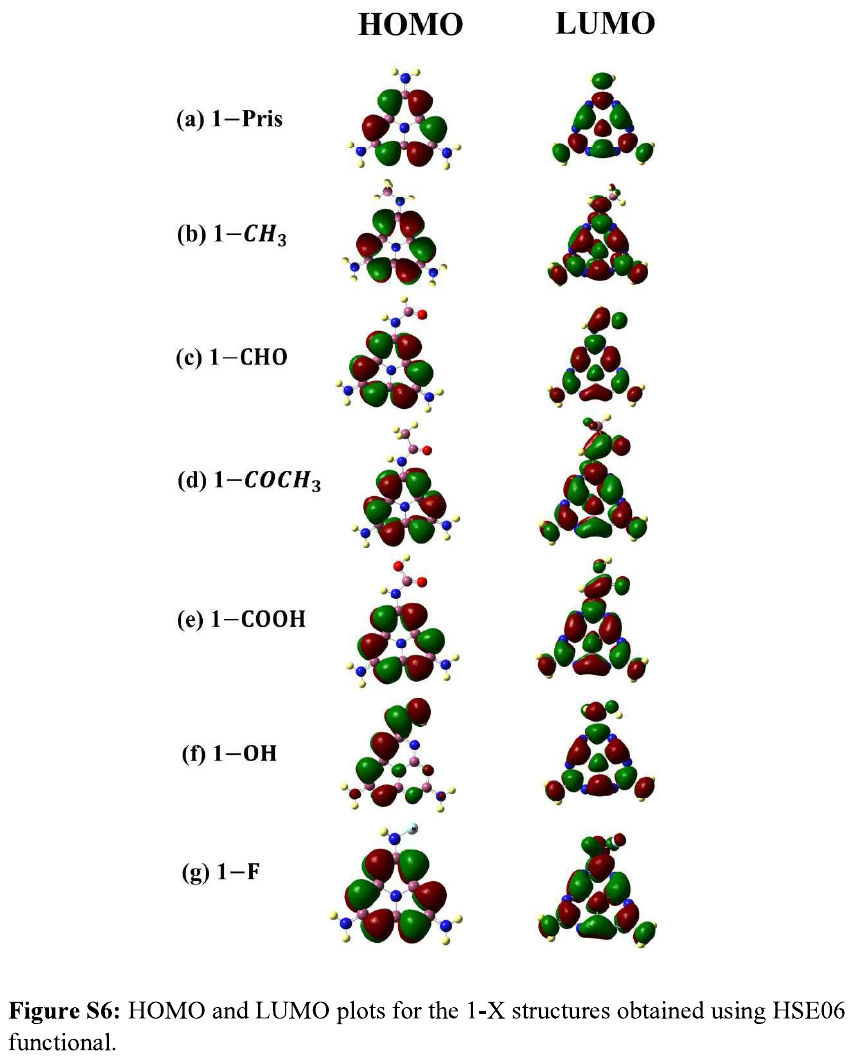}

\includegraphics[scale=0.85]{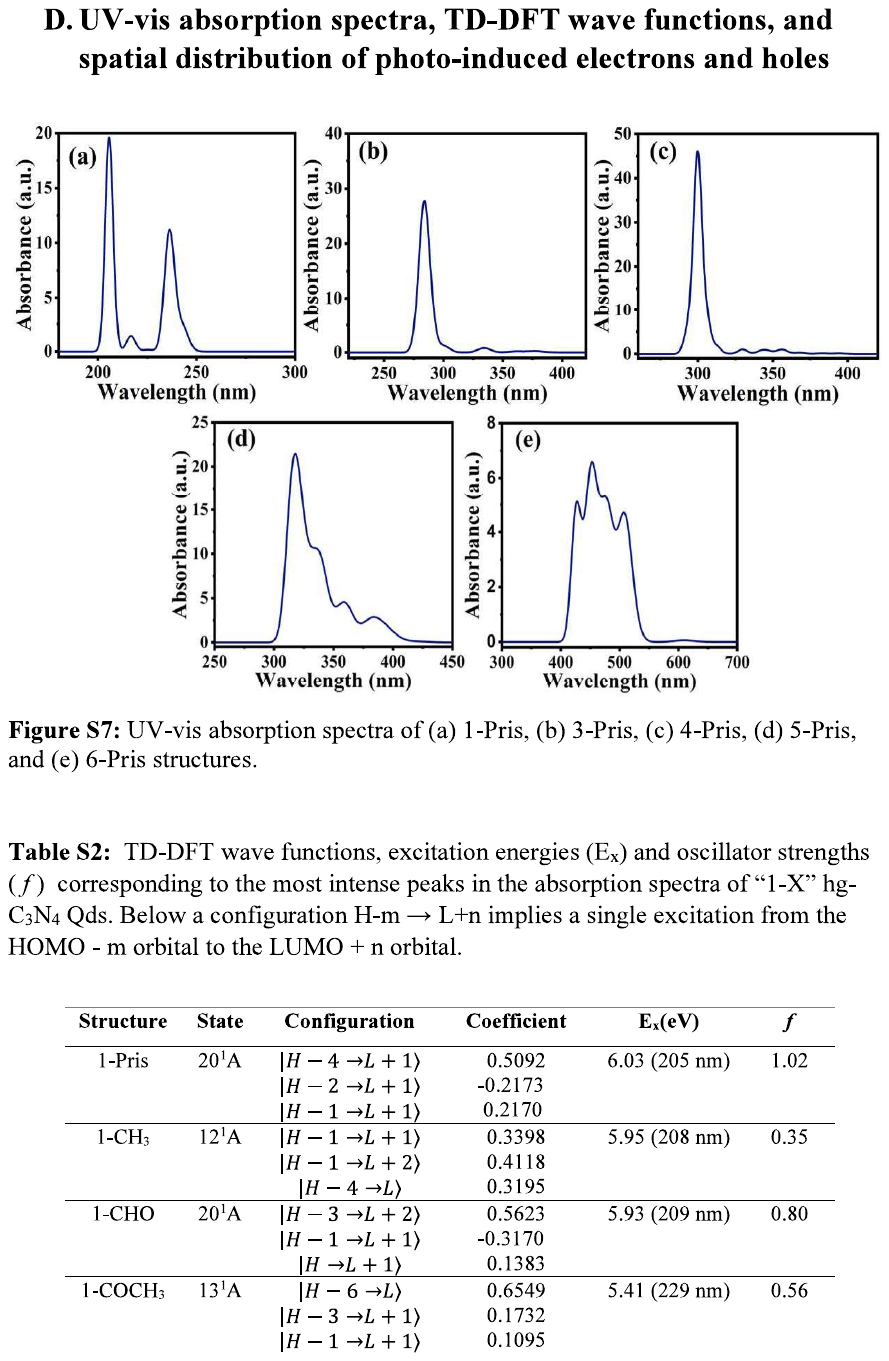}

\includegraphics[scale=0.85]{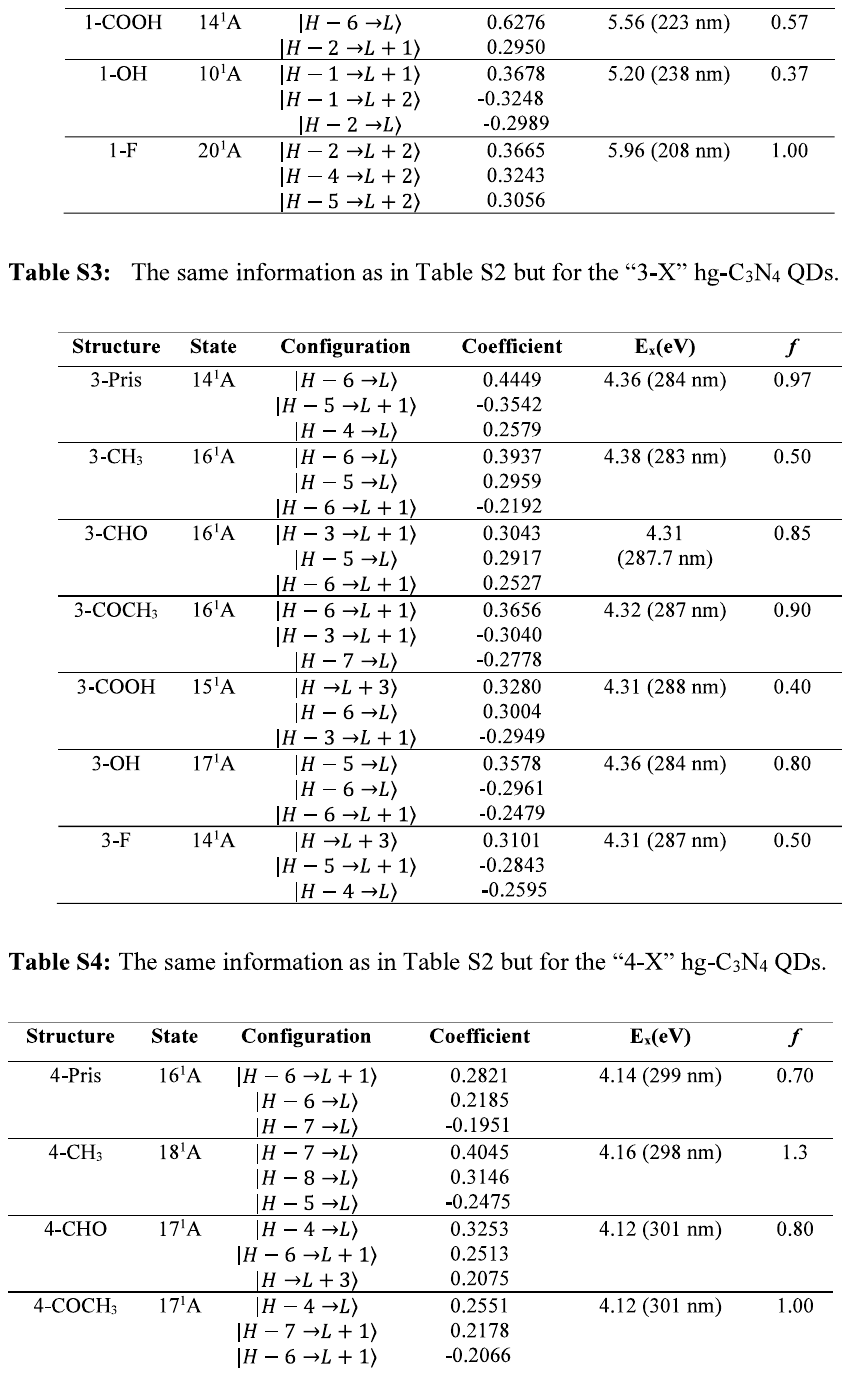}

\includegraphics[scale=0.85]{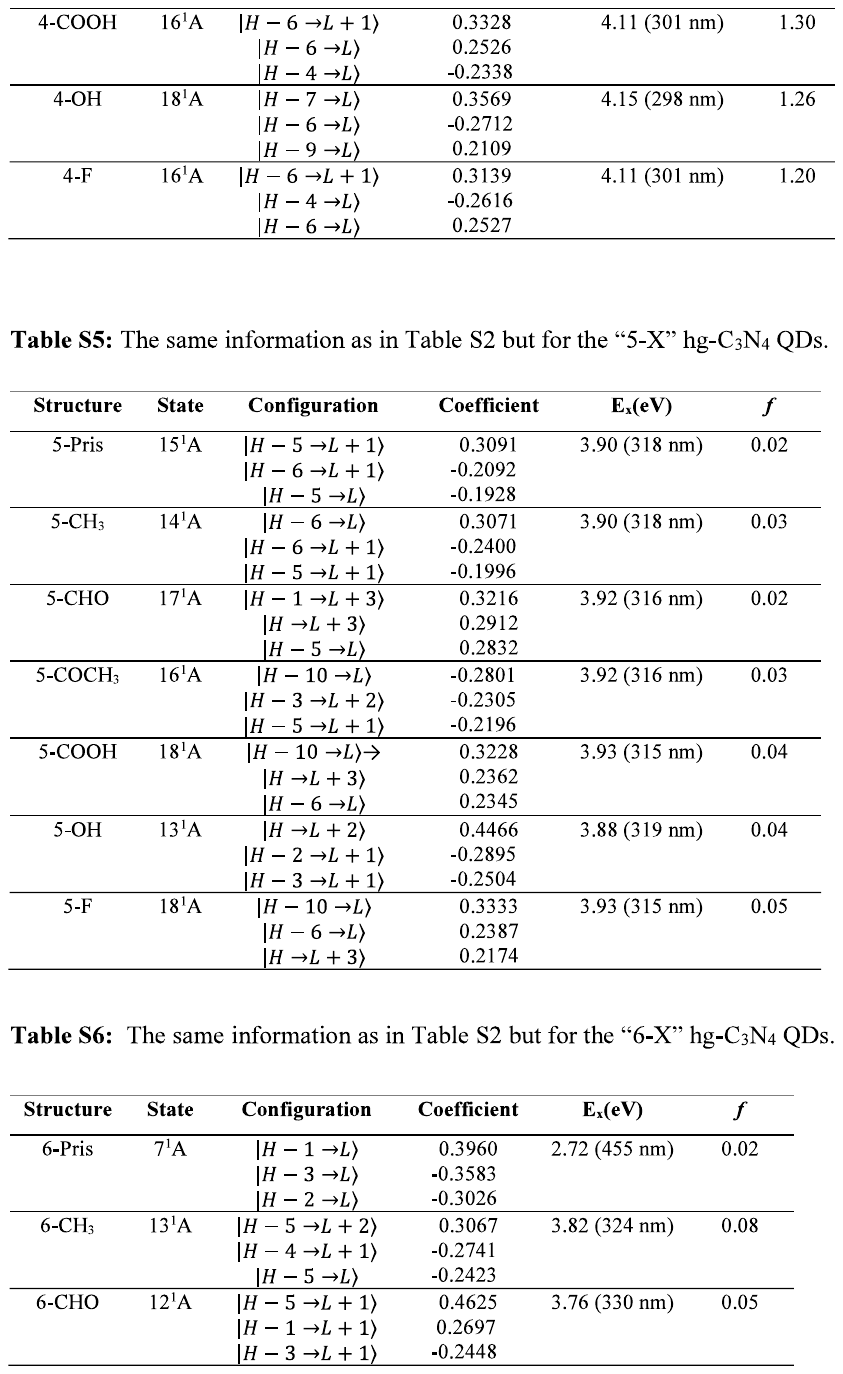}

\includegraphics[scale=0.85]{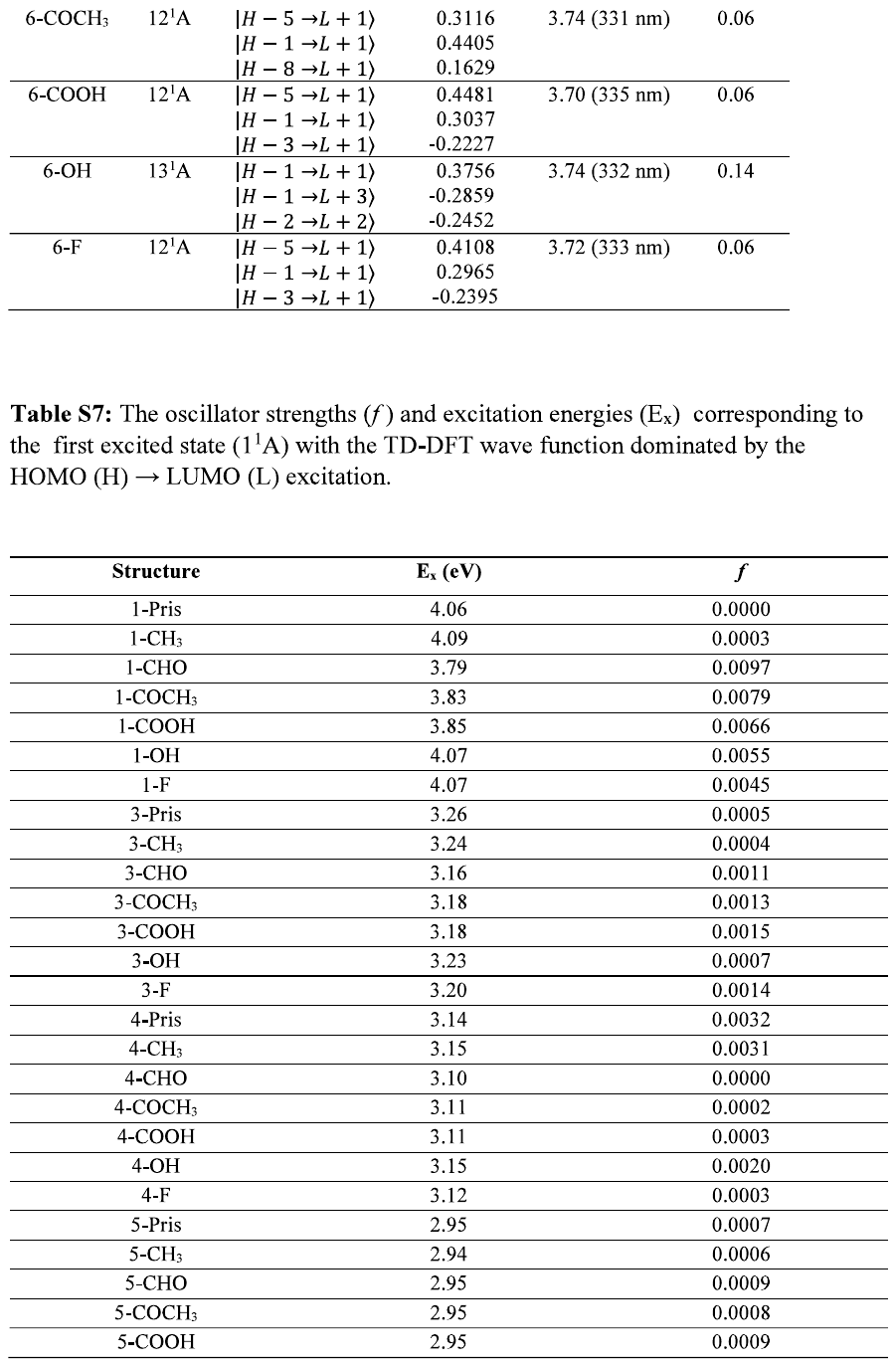}

\includegraphics[scale=0.85]{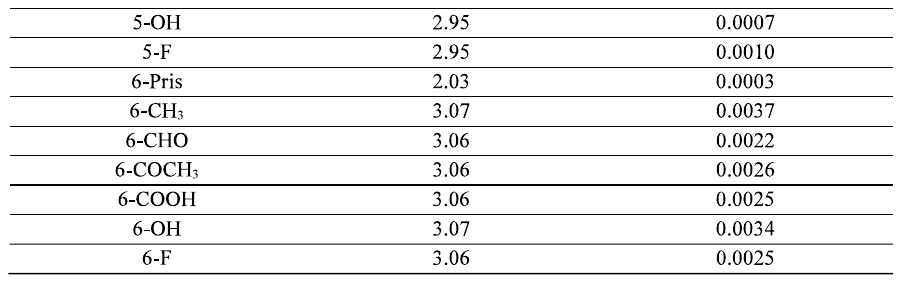}

\includegraphics[scale=0.85]{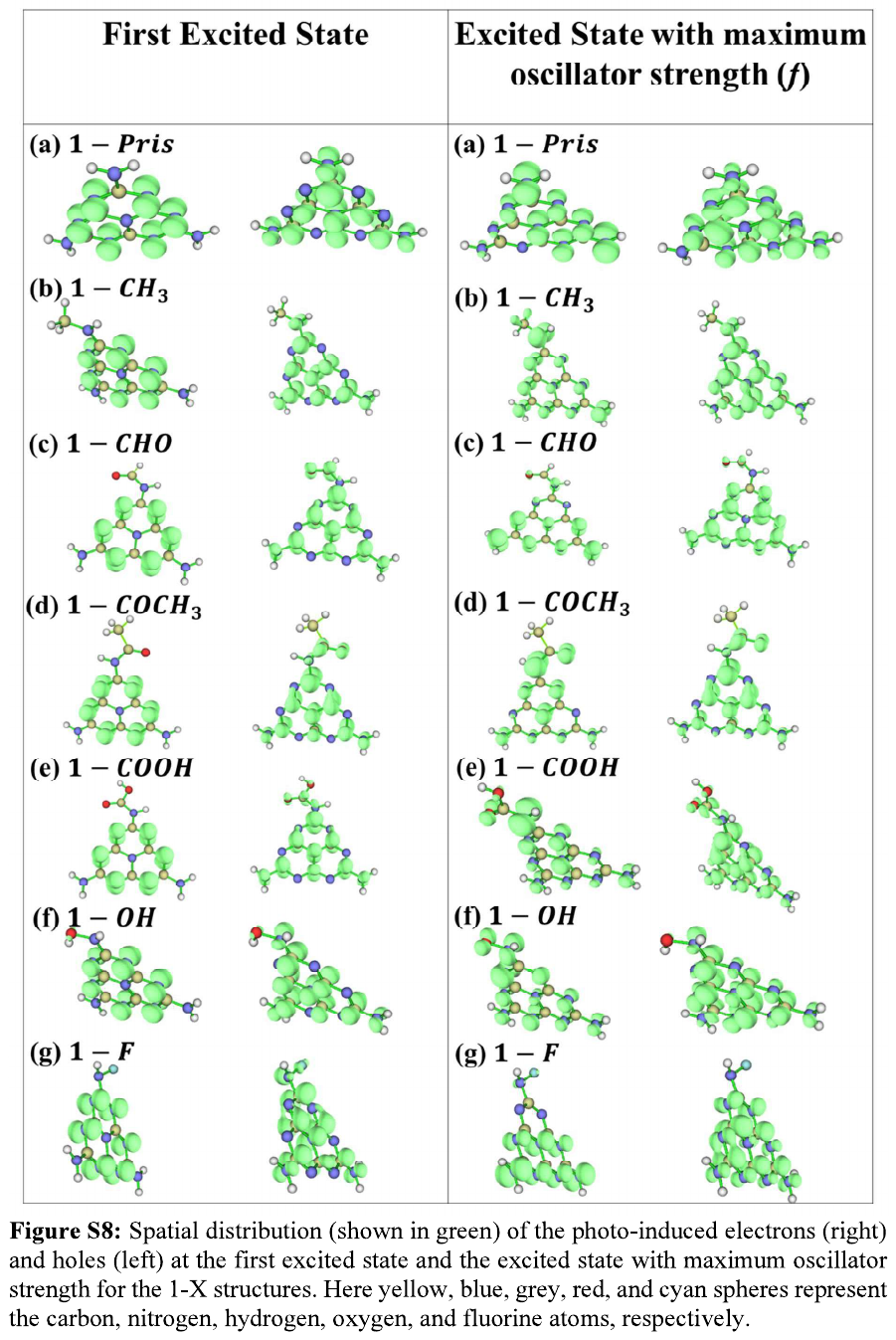}

\includegraphics[scale=0.85]{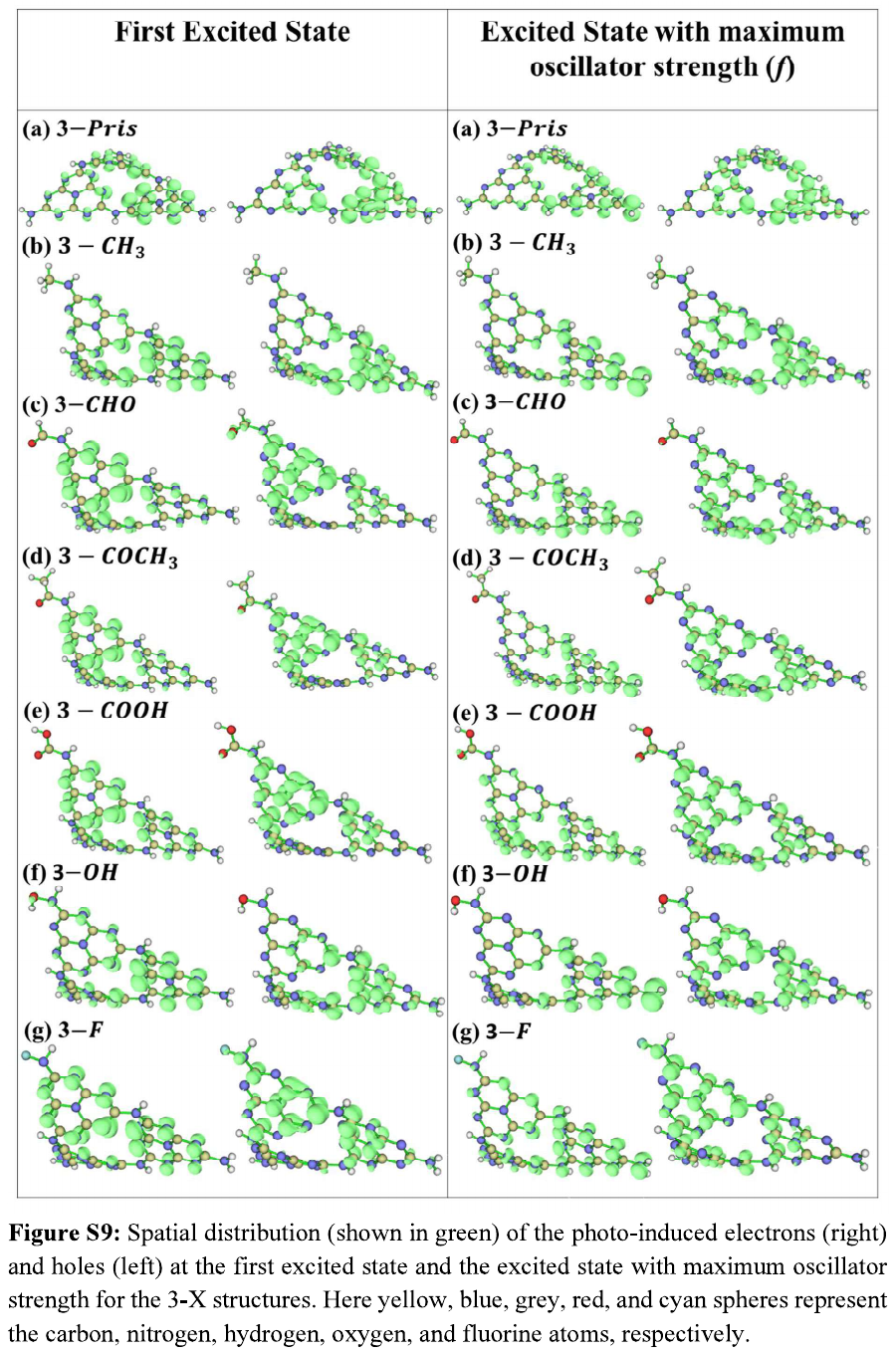}

\includegraphics[scale=0.85]{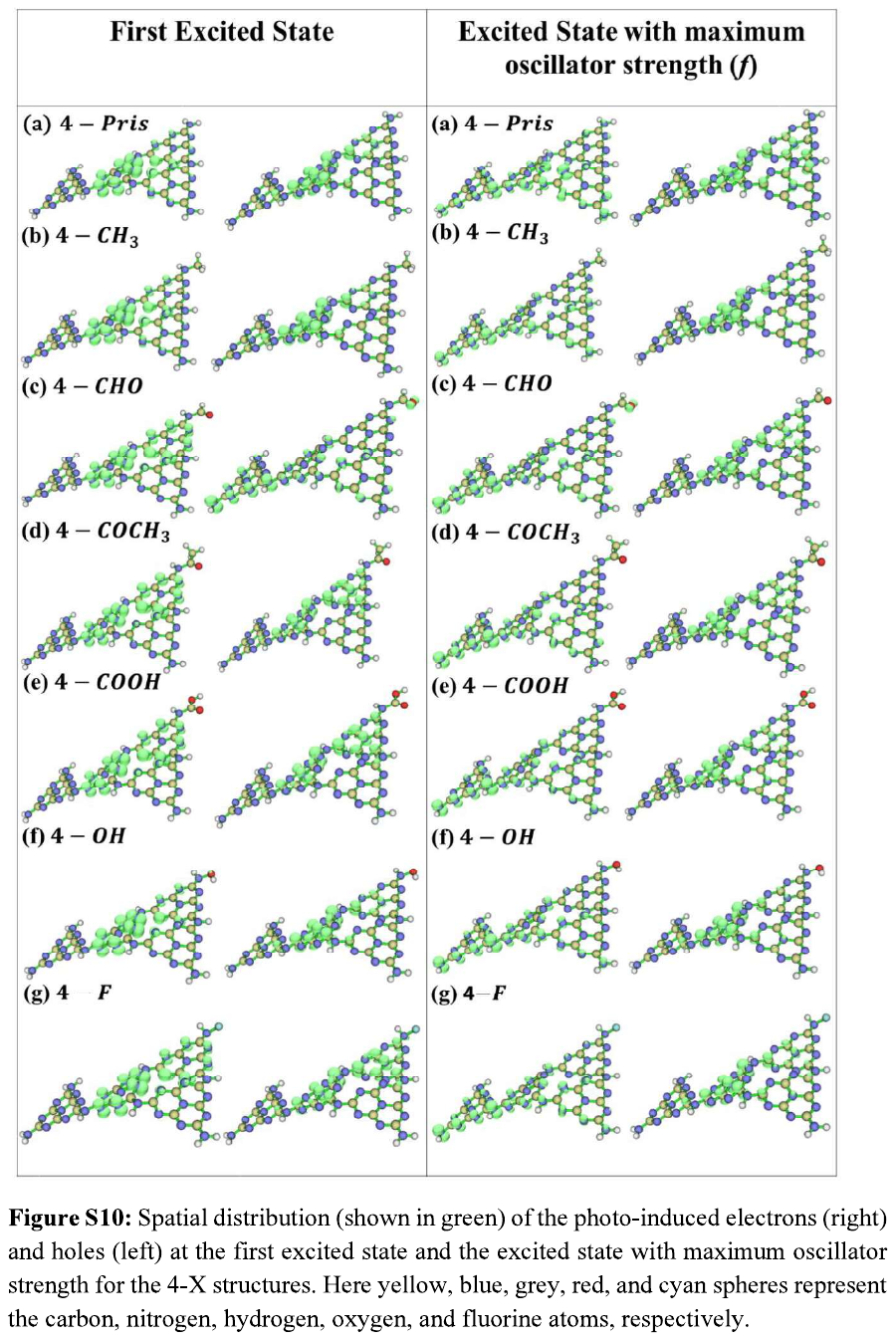}

\includegraphics[scale=0.85]{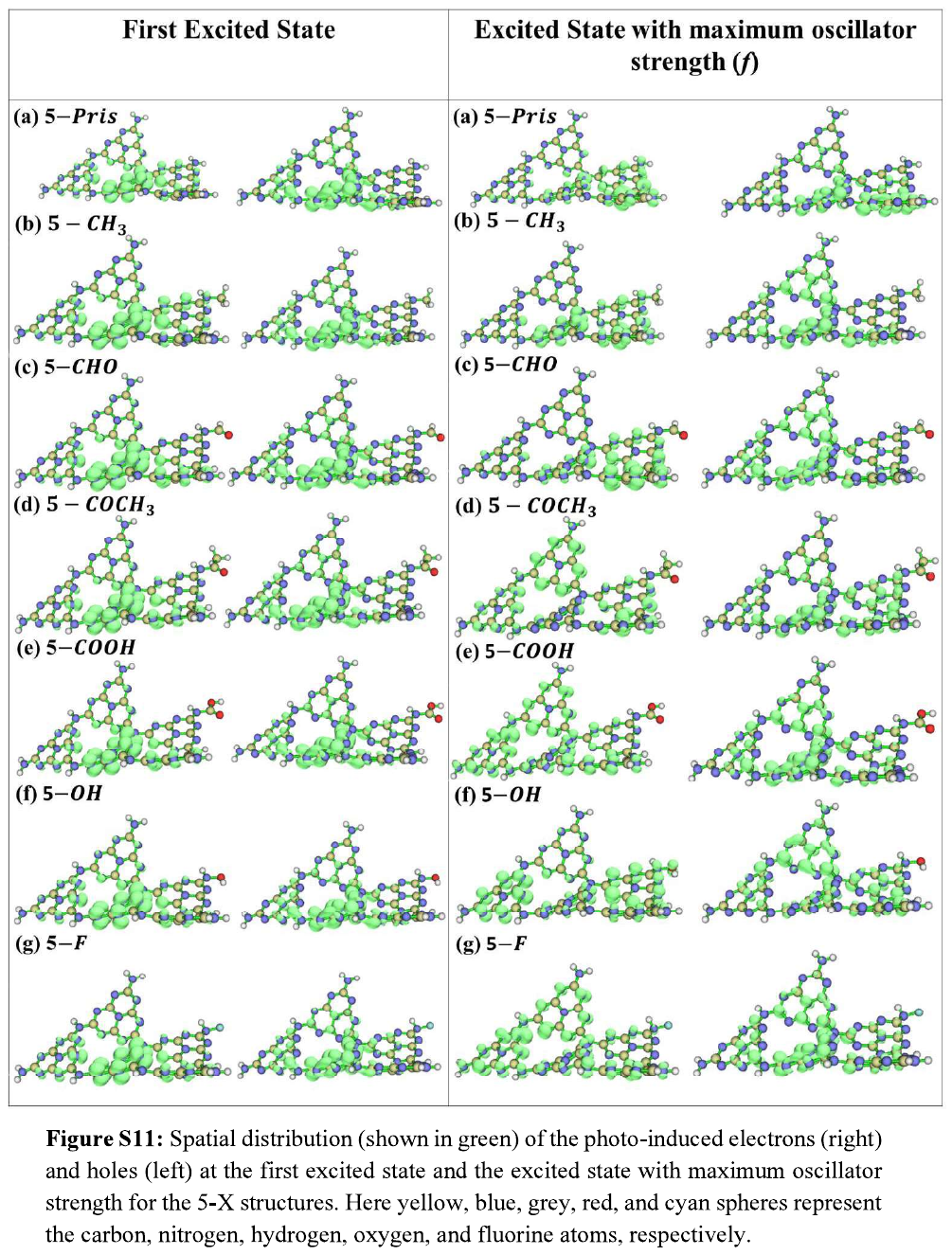}

\includegraphics[scale=0.85]{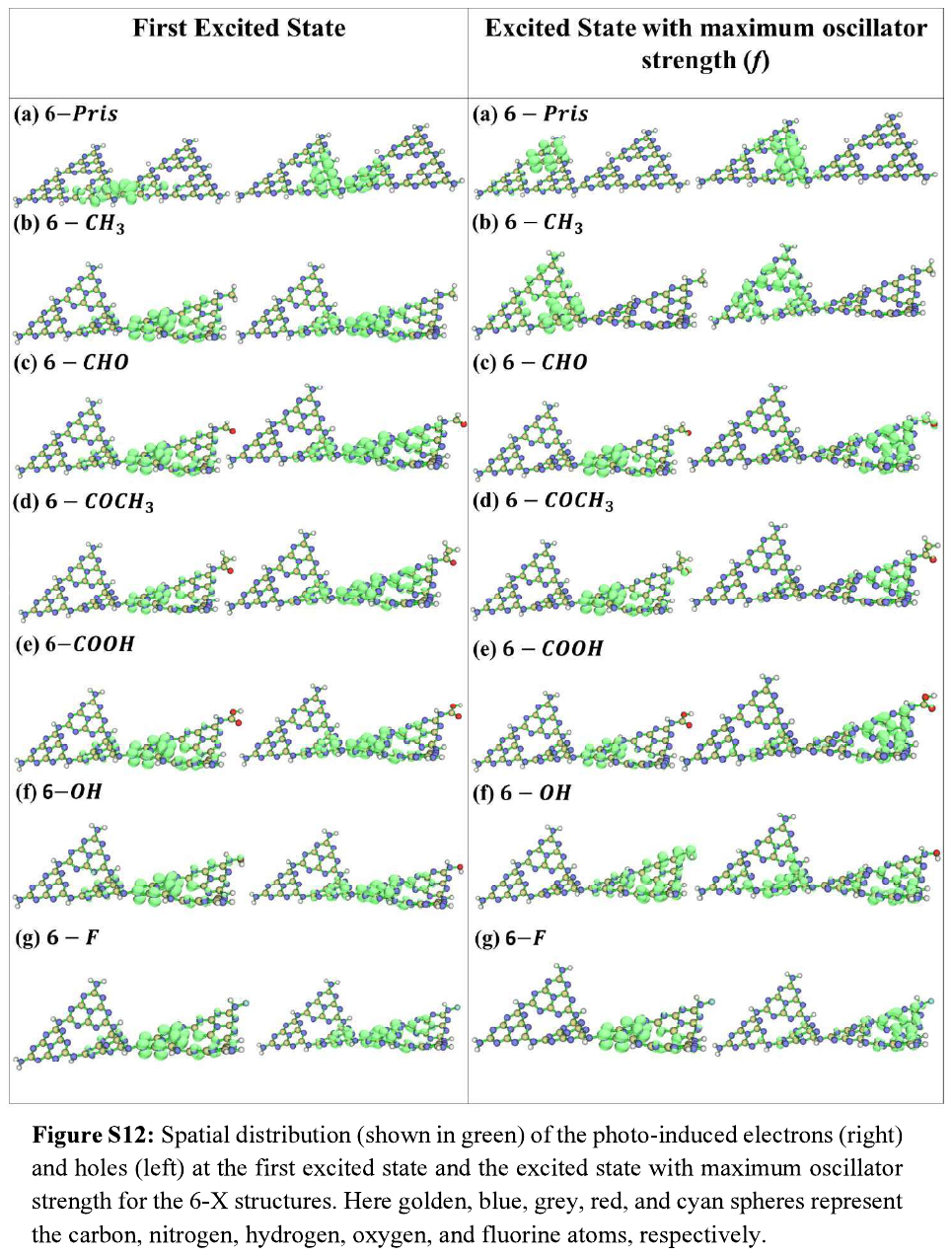}
\end{document}